\documentclass[]{spie}  

\usepackage{amsmath,amsfonts,amssymb}
\usepackage{graphicx}
\usepackage[colorlinks=true, allcolors=blue]{hyperref}
\usepackage{multirow}

\title{Design of SCALES: A 2-5 Micron Coronagraphic Integral Field Spectrograph for Keck Observatory}

\author[a]{Andrew Skemer}
\author[b]{R. Deno Stelter}
\author[c]{Stephanie Sallum}
\author[b]{Nicholas MacDonald}
\author[b]{Renate Kupke}
\author[b]{Christopher Ratliffe}
\author[d]{Ravinder Banyal}
\author[d]{Amirul Hasan}
\author[d]{Hari Mohan Varshney}
\author[e]{Arun Surya}
\author[d]{Ajin Prakash}
\author[d]{Sivarani Thirupathi}
\author[d]{Ramya Sethuraman}
\author[d]{Govinda K.V.}
\author[f]{Michael P. Fitzgerald}
\author[f]{Eric Wang}
\author[g]{Marc Kassis}
\author[h]{Olivier Absil}
\author[g]{Carlos Alvarez}
\author[i]{Natasha Batalha}
\author[j]{Marc-Andr\'e Boucher}
\author[k]{Cyril Bourgenot}
\author[l]{Timothy Brandt}
\author[a]{Zackery Briesemeister}
\author[m]{Katherine de Kleer}
\author[n]{Imke de Pater}
\author[b]{William Deich}
\author[d]{Devika Divakar}
\author[j]{Guillaume Filion}
\author[j]{\'Etienne Gauvin}
\author[a,b]{Michael Gonzales}
\author[i]{Thomas Greene}
\author[b]{Philip Hinz}
\author[a]{Rebecca Jensen-Clem}
\author[f]{Christopher Johnson}
\author[a]{Isabel Kain}
\author[b]{Gabriel Kruglikov}
\author[c]{Mackenzie Lach}
\author[j]{Jean-Thomas Landry}
\author[a]{Jialin Li}
\author[o]{Michael C. Liu}
\author[g]{James Lyke}
\author[f]{Kenneth Magnone}
\author[g]{Eduardo Marin}
\author[a]{Emily C. Martin}
\author[c]{Raquel A. Martinez}
\author[m]{Dimitri Mawet}
\author[a,c]{Brittany Miles}
\author[b]{Dale Sandford}
\author[p]{Patrick Sheehan}
\author[f]{Ji Man Sohn}
\author[q]{Jordan Stone}

\affil[a]{UC Santa Cruz, Santa Cruz, CA, USA}
\affil[b]{UC Observatories, Santa Cruz, CA, USA}
\affil[c]{UC Irvine, Irvine, CA, USA}
\affil[d]{Indian Institute of Astrophysics, Bengaluru, India}
\affil[e]{Tata Institute of Fundamental Research, Mumbai, India}
\affil[f]{UCLA, Los Angeles, CA, USA}
\affil[g]{Keck Observatory, Waimea, HI, USA}
\affil[h]{STAR Institute, Universit\'e de Li\`ege, Li\`ege, Belgium}
\affil[i]{NASA Ames Research Center, Moffett Field, CA, USA}
\affil[j]{OMP inc., Quebec City, QC, Canada}
\affil[k]{Durham Precision Optics, Durham, UK}
\affil[l]{UC Santa Barbara, Santa Barbara, CA, USA}
\affil[m]{California Institute of Technology, Pasadena, CA, USA}
\affil[n]{UC Berkeley, Berkeley, CA, USA}
\affil[o]{Institute for Astronomy, University of Hawai'i, Honolulu HI}
\affil[p]{Northwestern University, Evanston, IL, USA}
\affil[q]{United States Naval Research Laboratory, Washington, DC, USA}

\authorinfo{Further author information: (Send correspondence to Andrew Skemer: askemer@ucsc.edu)}

\begin{document} 
\maketitle

\begin{abstract}
We present the design of SCALES (Slicer Combined with Array of Lenslets for Exoplanet Spectroscopy) a new 2-5 micron coronagraphic integral field spectrograph under construction for Keck Observatory. SCALES enables low-resolution (R$\sim$50) spectroscopy, as well as medium-resolution (R$\sim$4,000) spectroscopy with the goal of discovering and characterizing cold exoplanets that are brightest in the thermal infrared. Additionally, SCALES has a 12x12" field-of-view imager that will be used for general adaptive optics science at Keck. We present SCALES's specifications, its science case, its overall design, and simulations of its expected performance. Additionally, we present progress on procuring, fabricating and testing long lead-time components.
\end{abstract}


\section{INTRODUCTION}
\label{sec:intro}  
Thanks to progress in adaptive optics and instrumentation, we have now directly detected a small number of exoplanets, with $\sim20-25$ imaged and a handful characterized with spectroscopy\cite{2022arXiv220505696C}.
In the last decade, coronagraphic integral field spectrographs, including GPI\cite{2014PNAS..11112661M}, SPHERE\cite{2008SPIE.7014E..18B}, and CHARIS\cite{2015SPIE.9605E..1CG}, have taken advantage of spatial and spectroscopic differences between planets and speckles to discover new exoplanets that were previously hidden in the glare of their host stars.  
All of these integral field spectrographs are coupled with extreme adaptive optics systems that are designed to provide the high-contrasts ($\sim10^6$) necessary to detect a faint exoplanet next to its bright host star.
Despite all of this progress, even with extreme adaptive optics systems, we can currently only access the brightest planets with the widest angular separations from their stars: young giant planets that are significantly hotter, more massive, and on wider orbits than Jupiter. 
Expanding the planet characterization parameter space requires novel instrumentation. 

Here we present the design of a new instrument, the Slicer Combined with Array of Lenslets for Exoplanets Spectroscopy (SCALES)\footnote{Previously, the instrument has been called Santa Cruz Array of Lenslets for Exoplanet Spectroscopy \cite{2018SPIE10702E..A5S,2020SPIE11447E..64S}.  However, given the broad partnership that has coalesced around this instrument concept, we have changed the name to Slicer Combined with Array of Lenslets for Exoplanet Spectroscopy, which has the same acronym: SCALES.  This also follows from Arizona Lenslets for Exoplanet Spectroscopy (ALES)\cite{2015SPIE.9605E..1DS,2018SPIE10702E..3FS}, which prototyped many of the technologies used in SCALES.}, which adopts many of the same technologies as this previous generation of high-contrast instruments (e.g., integral field spectroscopy, coronagraphy, extreme adaptive optics).  
However, while GPI, SPHERE and CHARIS all operate in the near-infrared (1-2 $\mu$m), SCALES is designed to operate in the thermal infrared (2-5 $\mu$m), where self-luminous exoplanets are brighter and easier to detect in the glare of their host star.  As a result, SCALES will be able to detect and characterize colder and lower-mass planets than were previously accessible to high-contrast exoplanet-imaging instruments.

\begin{figure}[h]
    \centering
    \includegraphics[width=0.6\textwidth]{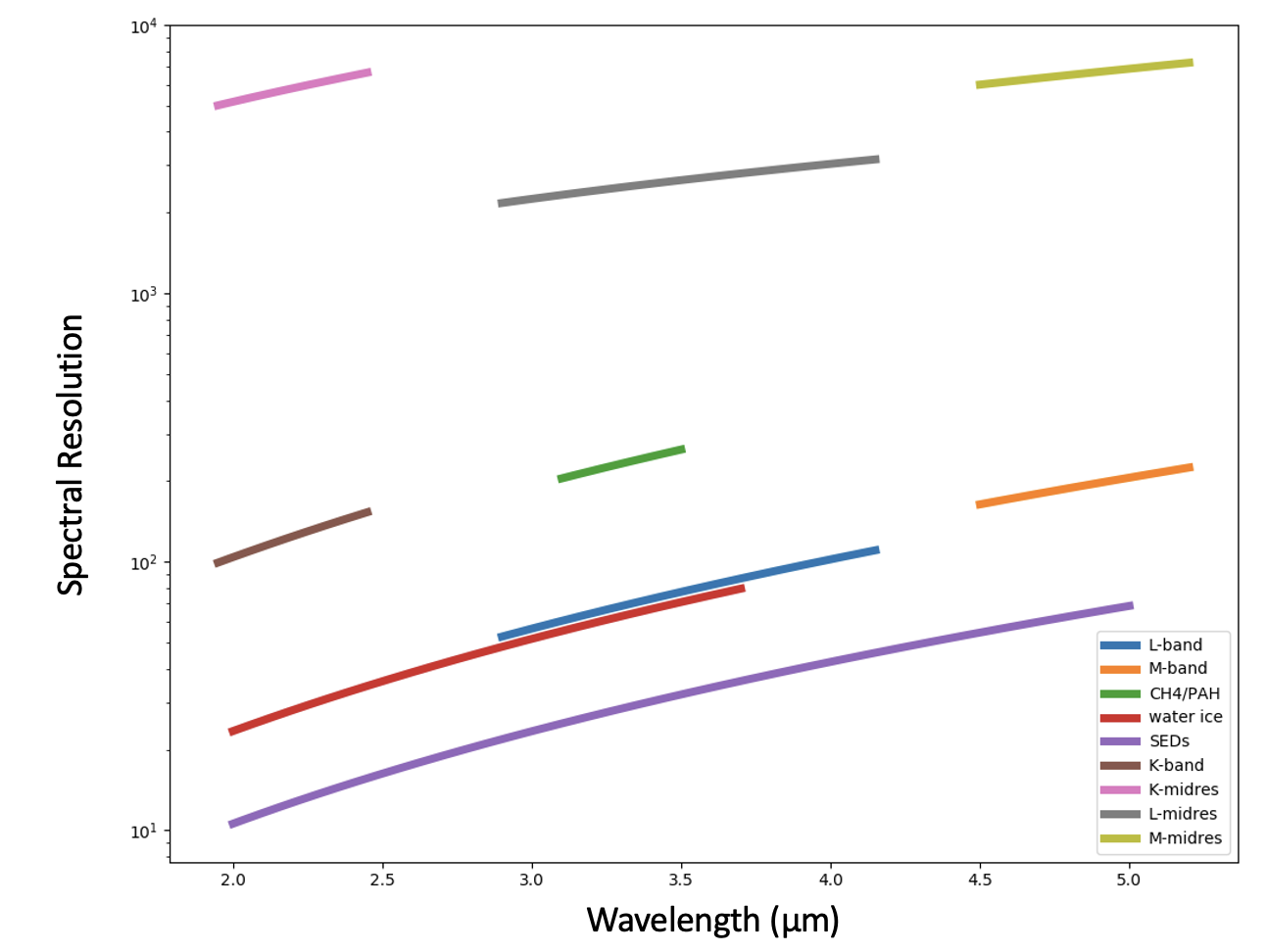}
    \caption{Spectral resolutions of the SCALES IFS modes
    }
    \label{fig:resolution}
\end{figure}

A summary of SCALES's 3 main observational modes is listed in Table \ref{SCALES specs}.  Spectral resolutions are shown in Figure \ref{fig:resolution}.

\begin{table}[ht]
\caption{SCALES Top-Level Specifications} 
\label{SCALES specs}
\begin{center}       
\begin{tabular}{|l|ll|ll|l|}
\hline
\multicolumn{1}{|c|}{\textbf{}}      & \multicolumn{2}{c|}{\textbf{Low-Resolution IFS}} & \multicolumn{2}{c|}{\textbf{Medium-Resolution IFS}}                                  & \multicolumn{1}{c|}{\textbf{Imager}}                                                            \\ \hline
\multirow{6}{*}{\textbf{Wavelength}} & \multicolumn{1}{l|}{2.0-2.4$\mu$m}  & R$\sim$150 & \multicolumn{1}{l|}{\multirow{2}{*}{2.0-2.4$\mu$m}}  & \multirow{2}{*}{R$\sim$6,000} & \multirow{6}{*}{\begin{tabular}[c]{@{}l@{}}Up to 16 filters \\ spanning 1-5$\mu$m\end{tabular}} \\ \cline{2-3}
                                     & \multicolumn{1}{l|}{2.0-4.0$\mu$m}  & R$\sim$50  & \multicolumn{1}{l|}{}                                &                               &                                                                                                 \\ \cline{2-5}
                                     & \multicolumn{1}{l|}{2.0-5.0$\mu$m}  & R$\sim$35  & \multicolumn{1}{l|}{\multirow{2}{*}{2.9-4.15$\mu$m}} & \multirow{2}{*}{R$\sim$3,000} &                                                                                                 \\ \cline{2-3}
                                     & \multicolumn{1}{l|}{2.9-4.15$\mu$m} & R$\sim$80  & \multicolumn{1}{l|}{}                                &                               &                                                                                                 \\ \cline{2-5}
                                     & \multicolumn{1}{l|}{3.1-3.5$\mu$m}  & R$\sim$200 & \multicolumn{1}{l|}{\multirow{2}{*}{4.5-5.2$\mu$m}}  & \multirow{2}{*}{R$\sim$7,000} &                                                                                                 \\ \cline{2-3}
                                     & \multicolumn{1}{l|}{4.5-5.2$\mu$m}  & R$\sim$200  & \multicolumn{1}{l|}{}                                &                               &                                                                                                 \\ \hline
\textbf{Field of View}               & \multicolumn{2}{l|}{2.15$\times$2.15"}           & \multicolumn{2}{l|}{0.36$\times$0.34"}                                               & 12.3$\times$12.3"                                                                               \\ \hline
\textbf{Spatial Sampling}            & \multicolumn{2}{l|}{0.02"}                       & \multicolumn{2}{l|}{0.02"}                                                           & 0.006"                                                                                          \\ \hline
\textbf{Coronagraphy}                & \multicolumn{2}{l|}{Vector-Vortex}               & \multicolumn{2}{l|}{Vector-Vortex}                                                   & TBD                                                                                             \\ \hline
\end{tabular}
\end{center}
\end{table}

For more information on SCALES, please see the following proceedings in this conference:

\noindent\textbf{Optical Design}: Paper No. 12184-159 (Renate Kupke et al.)\cite{Reni2022}\\
\textbf{Imaging Channel}: Paper No. 12188-65 (Ravinder Banyal et al.)\cite{Banyal2022}\\
\textbf{Slicer for Med-Res IFS}: Paper No. 12184-154 (Stelter et al.)\cite{DenoSlenslit2022}\\
\textbf{Cold-Stop / Lyot Stop}: Paper No. 12185-332 (Li et al.)\cite{Jialin2022}\\
\textbf{Aperture Masks}: Paper No. 12183-89 (Lach et al.)\cite{Lach2022}\\
\textbf{Keck Instrument Development}: Paper No. 12184-4 (Kassis et al.)\cite{Kassis2022}\\

\newpage
\section{SCIENCE OVERVIEW}

\begin{figure}[h]
    \centering
    \includegraphics[width=0.6\textwidth]{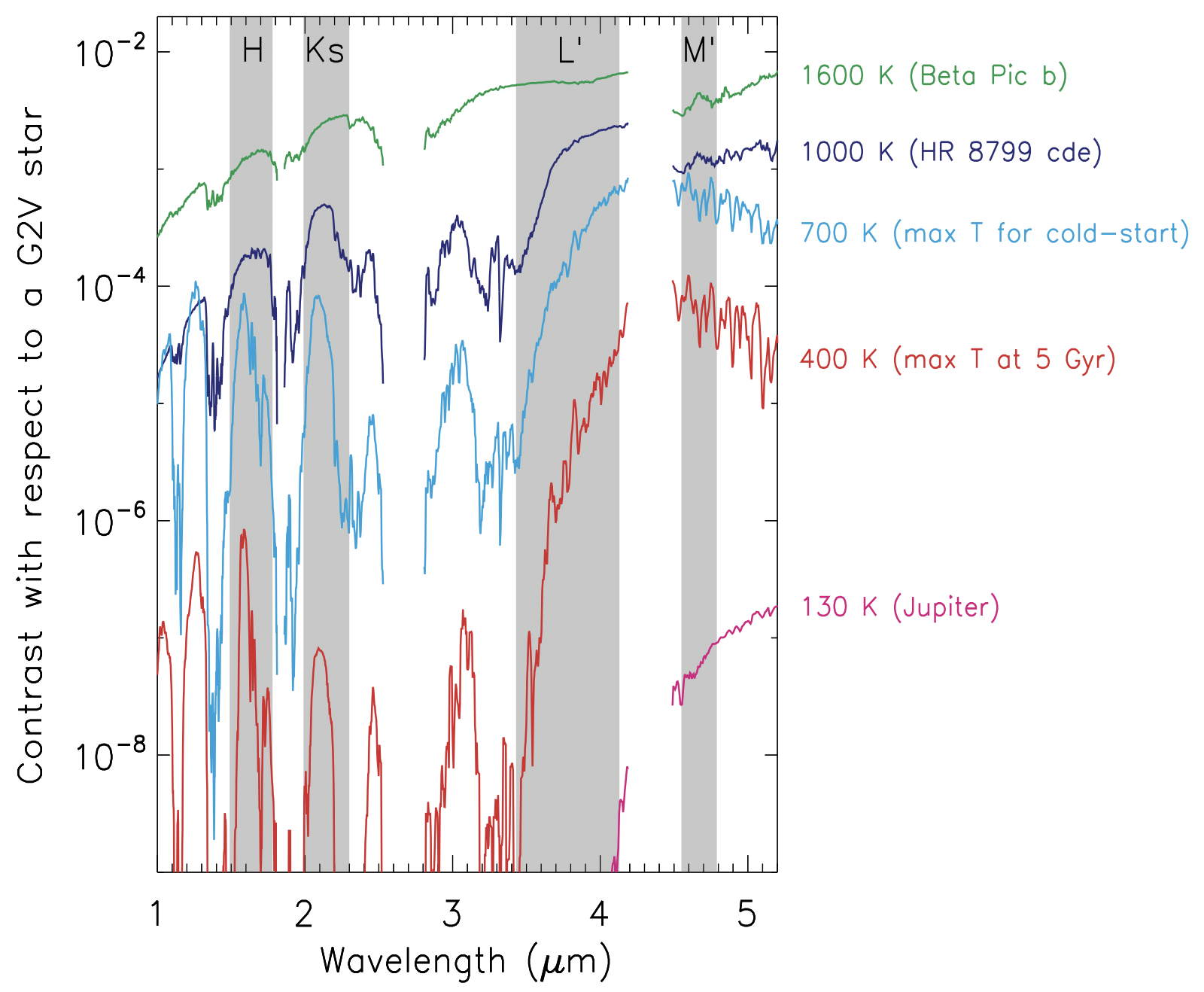}
    \caption{Characteristic exoplanet-to-star contrasts (flux ratios) vs. wavelength showing that giant exoplanets can be detected with lower contrast in the thermal-infrared (3-5 $\mu$m) than in the near-infrared (1-2 $\mu$m), and that this effect is most dramatic for cold planets. While most directly-imaged planets to date are relatively warm ($>$ 700 K), the majority of self-luminous exoplanets are much cooler. Figure reprinted\cite{2014ApJ...792...17S}.
    }
    \label{fig:contrast}
\end{figure}

\begin{figure}[h]
    \centering
    \includegraphics[width=1.0\textwidth]{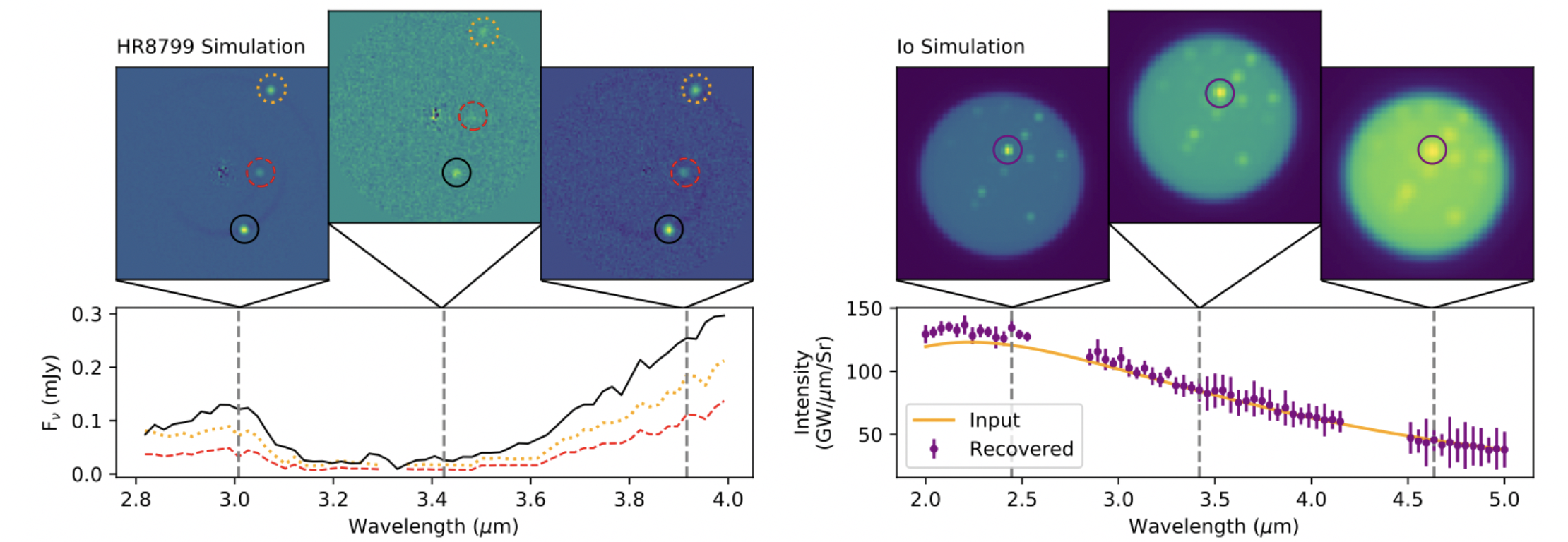}
    \caption{Simulated SCALES data cubes and spectra of the HR 8799 planetary system (left) and Io (right), generated with SCALES’ end-to-end simulator \cite{2020SPIE11447E..4ZB}. This takes into account astrophysical, environmental, and instrumental noise sources, including shot noise from the source, sky transmission and emission, SCALES’ optic-by-optic throughput, and detector systematics. We simulated raw science frames, and then reduced the simulated data using the ALES pipeline\cite{2018SPIE10702E..2QB} modified for SCALES. Note that HR 8799 has 4 planets, but when the star is centered in the SCALES field-of-view, only the inner 3 are visible. For the Io spectrum, the difference between the simulated blackbody and the reduced data is because the volcano is marginally resolved spatially.
    }
    \label{fig:sci-sim}
\end{figure}

SCALES combines the two most powerful methods for imaging exoplanets: (1) thermal infrared ($2-5 \mu$m) imaging, which detects exoplanets at wavelengths where they are bright (Figure \ref{fig:contrast}), and (2) integral-field spectroscopy, which distinguishes exoplanets from residual starlight based on the shapes of their spectral energy distributions. For a 300 K planet, this combination creates a $\sim4-5$ magnitude boost in sensitivity compared to an H-band IFS, and a $\sim$2.2 magnitude boost in sensitivity compared to an L-band imager\footnote{For a given planet temperature, we estimate colors in common bandpasses (imaging) or optimal bandpasses (IFS), using model atmospheres\cite{2014ApJ...787...78M}, and calculate contrasts with respect to a Raleigh-Jeans tail. The optimal filter provides an effective contrast boost for IFS data\cite{2018SPIE10702E..A5S}. For IFS data, we also independently adopt a 1 magnitude gain consistent with empirically-demonstrated IFS-based starlight speckle suppression \cite{2014SPIE.9148E..0UM,2015MNRAS.454..129V,SPHERE_manual}}. By operating at longer wavelengths than other high-contrast integral-field spectrographs, SCALES will extend the wavelength range we use to characterize planets, and also discover new planets (in particular, cold planets) that are not detectable with near-infrared instruments. Despite the competitiveness of the exoplanet imaging field, SCALES’ unique parameter space ensures that it will lead a broad range of new science.  End-to-end simulations of an exoplanet imaging observation and a Solar System synoptic program are shown in Figure \ref{fig:sci-sim}.

\subsection{SCALES Exoplanet Discovery}
SCALES will detect a wide range of exoplanets, including many beyond the limits of current instruments. Here, we discuss SCALES’ anticipated impact on exoplanet searches, focusing on detections not previously possible in large numbers (planets with known masses, cold-start planets, accreting protoplanets). These predictions were made using an end-to-end simulator accounting for SCALES’ optical design, as well as instrumental and atmospheric transmission, emission, and noise sources\cite{2020SPIE11447E..4ZB}.

\subsubsection{Old, Cold Gaia Exoplanets}
\begin{figure}[h]
    \centering
    \includegraphics[width=0.985\textwidth]{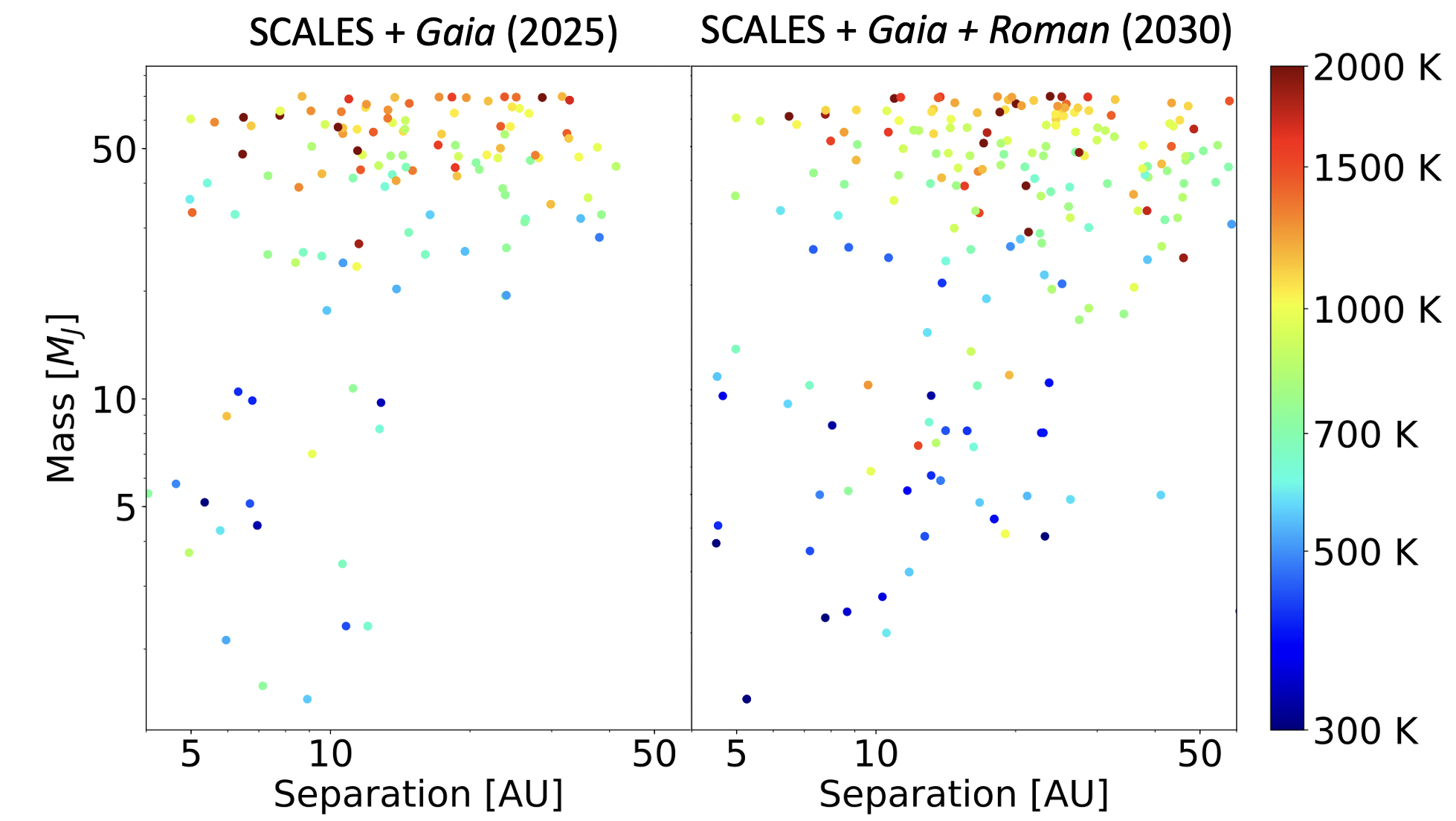}
    \caption{Simulated population of planets that would be discovered by the Gaia mission (left) and Gaia+Roman (right) that would also be detectable by SCALES given Keck’s anticipated adaptive optics performance in 2025. Almost all of the bona fide planets, both detected by SCALES and in the underlying population (M $<$ 13 MJ) are colder than what is currently the coldest directly-imaged planet ($\sim$700 K; \cite{2015Sci...350...64M}). Furthermore, the vast majority are located $<$ 0.5" from their host stars, highlighting the importance of high contrast at small inner working angle. These exoplanets will each have known masses, which is new for direct imaging, and will provide benchmarks for atmospheric and evolutionary studies. (We assume the WFIRST-CGI planet population models and SCALES contrast curves \cite{2019AJ....158..140B}. 
    }
    \label{fig:Gaia}
\end{figure}

Roughly when SCALES is completed, Gaia’s extended mission will be releasing a catalog of $\sim$70,000 exoplanets \cite{2014ApJ...797...14P}. Some of these will orbit nearby stars widely enough to be directly imaged. Using the SCALES predicted contrast curve  and following \cite{2019AJ....158..140B}, we predict 23 planets ($<$13 M$_{j}$) and 110 brown dwarfs ($>$13 M$_{j}$) will be detectable by SCALES at this time. Adding astrometry from \textit{Roman} in 2030, this increases to 30 planets and 176 brown dwarfs (Figure \ref{fig:Gaia}).  This is particularly exciting, since in general direct imaging cannot measure planet mass, and atmospheric properties are degenerate without masses. Furthermore, many of the new Gaia planets amenable to direct imaging will be $\sim$300 K (cold enough for water clouds and different atmospheric chemistry – e.g. NH$_{3}$, more CH$_{4}$ – to be visible in M-band spectra; \cite{2016ApJ...826L..17S}). These cannot be imaged with today’s near-infrared high-contrast IFSs, but can be imaged at longer wavelengths (Figure \ref{fig:contrast}).  These yields assume a Keck AO upgrade (a high-order deformable mirror) expected before SCALES commissioning. There is a risk that this upgrade may be delayed; however, even with current Keck AO, SCALES would still detect 6 Gaia planets in 2025 and 19 in 2030, nearly all of which would be colder than the current coldest directly imaged exoplanet ($\sim$700 K; \cite{2015Sci...350...64M}). Despite smaller yields in this worst-case scenario, each detection would be a high-impact result. More near-future AO upgrades are under consideration, including predictive control (a software upgrade progressing with promising results; \cite{2019SPIE11117E..0WJ,2022JATIS...8b9006V}), and an adaptive secondary mirror\cite{Hinz2022}.

\subsubsection{Directly Detecting Accreting Protoplanets in Gapped Protoplanetary Disks}
\begin{figure}[h]
    \centering
    \includegraphics[width=0.985\textwidth]{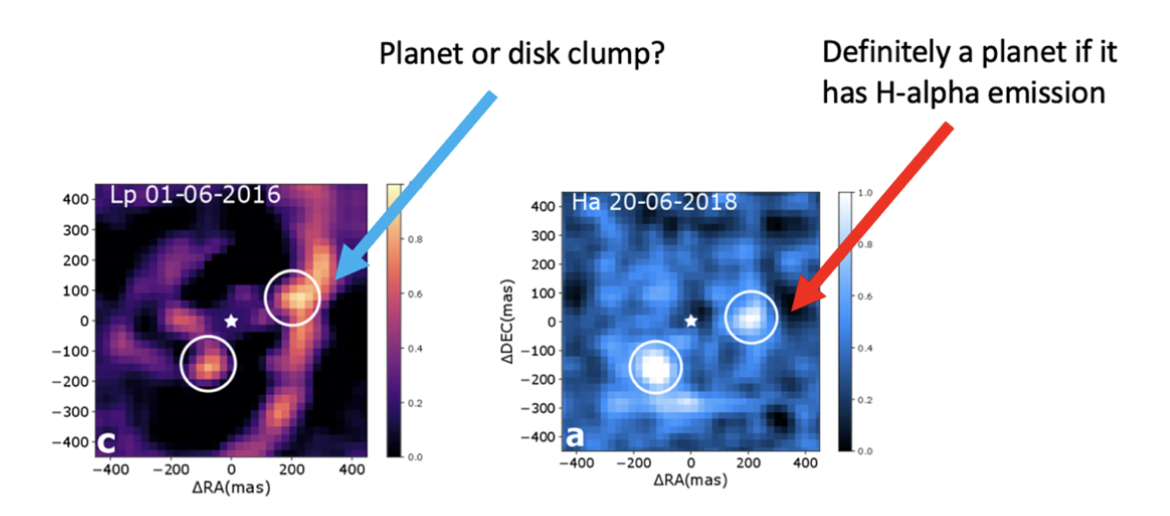}
    \caption{In imaging data, it is often impossible to differentiate between planets and circumstellar disk clumps. However, integral field spectroscopy data can distinguish the two scenarios via differences in their continuum slopes and via emission lines.  Image reprinted\cite{2019NatAs...3..749H}. 
    }
    \label{fig:PDS 70}
\end{figure}

By observing young planets in their nascent disks, it is possible to observe mass accretion directly. To date, only a couple of accreting exoplanets and candidates have been discovered \cite{2012ApJ...745....5K,2018A&A...617A..44K}. Part of the difficulty of imaging protoplanets is that they are usually embedded in protoplanetary disks, whose scattered light can resemble exoplanets \cite{2016SPIE.9907E..0DS}. Integral-field spectroscopy, specifically at the wavelengths where these planets peak in brightness (4 microns\cite{2015Natur.527..342S}), will distinguish between circumstellar disk scattered light and protoplanet emission by constraining spectral slopes and excesses in IR Hydrogen lines such as Br-$\gamma$ and Br-$\alpha$ (Figure \ref{fig:PDS 70}). Recent Keck AO upgrades are also aimed at imaging planets around young stars\cite{2018SPIE10703E..1ZB}.

\subsection{SCALES Exoplanet Characterization}
\begin{figure}[h]
    \centering
    \includegraphics[width=0.9\textwidth]{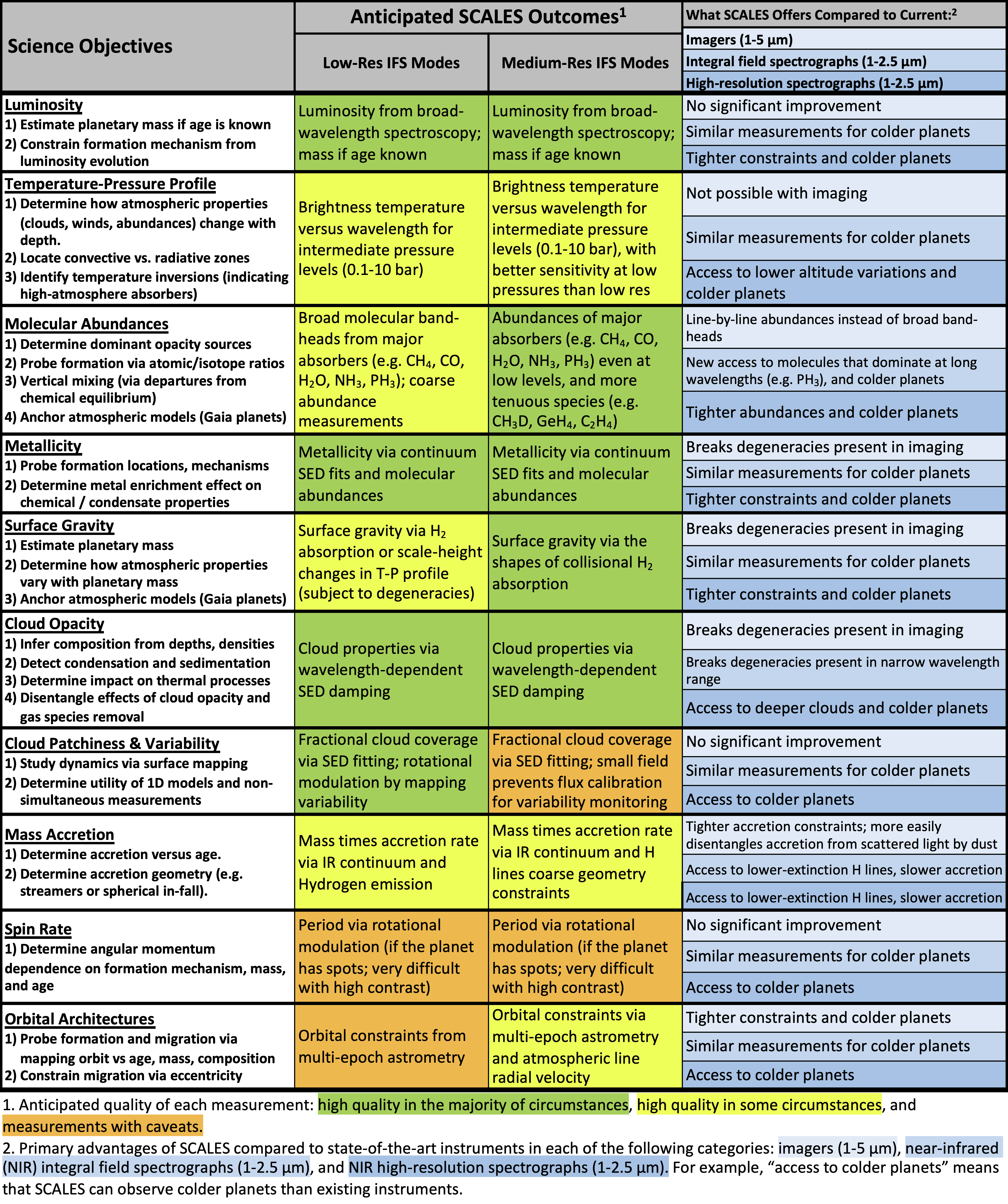}
    \caption{Anticipated SCALES science outcomes compared to current imagers, IFSs, and high-resolution spectrographs.  \label{fig:characterization}}
\end{figure}

Studying exoplanets beyond their locations, masses, and radii requires photometry and spectroscopy of the planets themselves. Spectroscopy can reveal a planet’s thermal structure, chemical composition, cloud properties, spatial inhomogeneity, and more\cite{2011ApJ...733...65B,2012ApJ...754..135M}. The observational signatures of these properties are often degenerate and require broad wavelength coverage to disentangle \cite{2014ApJ...792...17S,2016ApJ...817..166S}. Clouds, which are ubiquitous on exoplanets \cite{2013cctp.book..367M}, have optical properties that vary slowly with wavelength, and can be confused with thermal structure effects over a narrow bandpass \cite{2016ApJ...820...78L}. Molecules can probe chemical reactions, vertical mixing, and atomic abundances such as C/O ratios \cite{2013Sci...339.1398K}, but only if a relatively complete set are measured over the broad wavelength range where they are individually detectable \cite{2015ApJ...804...61B}.
SCALES will enable these studies by complementing existing near-infrared IFSs. In Figure \ref{fig:characterization}, we summarize its large range of exoplanet characterization opportunities. SCALES will break degeneracies that plague atmospheric characterization with imaging alone, resulting in detailed measurements of molecular abundances (including previously-unobserved species), metallicities, and surface gravities, all for a new planet population.  High-contrast medium-resolution spectroscopy is a new capability that SCALES will offer for the first time at any wavelength, allowing line-by-line identification of molecules, while simultaneously providing broad bandpass continuum measurements, all within the environment of a coronagraphic IFS.

\subsection{Additional Science Opportunities}
Thermal infrared imaging and spectroscopy are used for a wide range of Solar System, galactic, and extragalactic observations. Keck’s current thermal infrared imager, NIRC2, enables $\sim$70 papers per year.  SCALES will give users additional flexibility by allowing users to use a NIRC2-like imager alongside an integral field spectrograph.  A particular goal for SCALES is to enable a coordinated program of synoptic observations of Solar System objects, taking advantage of morning twilight, when regular observations are complete, but the sky brightness has not yet changed in the thermal infrared. 

\newpage
\section{INTEGRAL FIELD SPECTROGRAPH ARCHITECTURE}
SCALES features a lenslet-based integral field spectrograph as well as a hybrid lenslet/slicer integral field spectrograph (dubbed ``slenslit"), which share a collimator, camera and detector.  The lenslet-based integral field spectrograph produces low spectral resolutions, while the slenslit produces medium spectral resolutions over a smaller field-of-view.

The SCALES low-resolution IFS uses the lenslet-array architecture\cite{1995A&AS..113..347B}.  In this concept, each lenslet serves as a spatial pixel (or ``spaxel''), which samples the field and is dispersed into a spectrum by a downstream spectrograph.  The spectra are interleaved by rotating the disperser with respect to the lenslet array.  For exoplanet imaging, lenslet-based IFS’s are preferable to slicer IFS’s because the lenslet array samples the field before any optical aberrations are imparted by downstream spectroscopic optics.

\begin{figure}[hpt]
    \centering
    \includegraphics[width=0.7\textwidth]{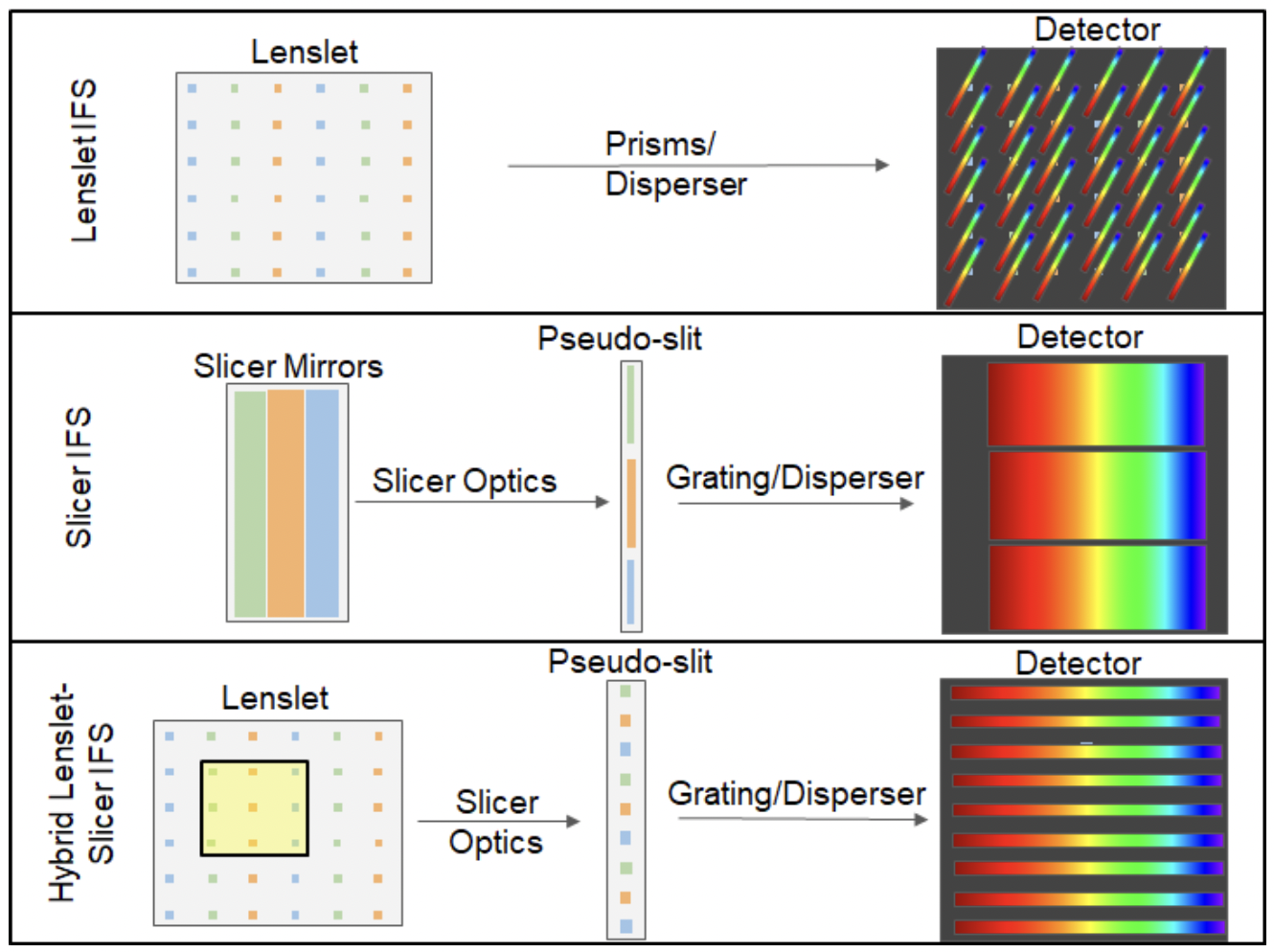}
    \caption[Illustration of IFS types.]
    {
        Illustration of the lenslet and slicer IFS formats, along with the hybrid lenslet-slicer (or ``slenslit'') concept used for the SCALES medium-resolution IFS.  The lenslet IFS is used by most exoplanet-imaging IFS’s because it samples the field before the downstream spectrograph imparts optical aberrations.  The slicer IFS has the benefit of creating longer spectra, but imparts optical aberrations.  The slenslit IFS combines the benefits of the lenslet IFS and the slicer IFS.
    }
    \label{fig-sp:IFS-types}
\end{figure}

Image slicers are an alternative IFS architecture that produce higher-resolution spectra than lenslet arrays.  However, their optics impart aberrations that are detrimental to high-contrast imaging.  For SCALES, we are proposing a new approach, which we call a slenslit, that combines a lenslet array with a slicer to achieve the best of both worlds: the lenslet array samples the field before the slicer and spectrograph impart optical aberrations, and the slicer re-formats the lenslet spots into a pseudo-slit that can be dispersed into longer spectra.  
An illustration of the concept is shown in Figure~\ref{fig-sp:IFS-types}.

Instrument stability is important for exoplanet imaging, so the SCALES fore-optics do not employ swappable magnifiers or lenslet arrays.
Because of this, and the shared requirements for wavelength, sampling, and crosstalk, the medium-resolution IFS uses the same lenslet pitch as the low-resolution IFS, and the spectra are separated by the same number of pixels at the detector.
Spectra are dispersed the length of the detector maximize spectral resolution and to allow use of 1st-order gratings which avoid spectroscopic order overlap.
For an H2RG, we can fit 340 spectra across the detector when each spectrum is separated from its neighbor by 6 pixels (or 108$\mu$m).  
This corresponds to an 18$\times$18 lenslet field-of-view (after a full optical design, we are only able to fit a 17$\times$18 lenslet field-of-view).  
The lenslet pitch is 341$\mu$m, which is close to 3 times 108$\mu$m (the separation of the low-resolution spectra). 
Therefore, a slicer that interleaves by 3 will achieve the same 6-pixel spectrum separation for the medium-resolution mode as the low-resolution mode.  
An illustration of the interleaving is shown in Figure~\ref{fig-sp:IFS-types}.  

A summary of the IFS properties and a plot of its spectral resolutions is shown in Table~\ref{tab-sp:ifs-summary}.

\begin{table}[]
\begin{center}       
\caption{Summary of IFS Layout}
\label{tab-sp:ifs-summary}
\begin{tabular}{|l|ll|ll|}
\hline
\multicolumn{1}{|c|}{\textbf{}}     & \multicolumn{2}{c|}{\textbf{Low-Resolution IFS}} & \multicolumn{2}{c|}{\textbf{Medium-Resolution IFS}} \\ \hline
\textbf{IFS Type}                   & \multicolumn{2}{l|}{Lenslet}                     & \multicolumn{2}{l|}{Slenslit}                       \\ \hline
\textbf{Number of Spaxels}          & \multicolumn{2}{l|}{108x108}                     & \multicolumn{2}{l|}{17x18}                          \\ \hline
\textbf{Plate Scale}                & \multicolumn{2}{l|}{0.02"}                       & \multicolumn{2}{l|}{0.02"}                          \\ \hline
\textbf{Field-of-View}              & \multicolumn{2}{l|}{2.14"$\times$2.14"}          & \multicolumn{2}{l|}{0.34"$\times$0.36"}             \\ \hline
\textbf{Length of Spectra}          & \multicolumn{2}{l|}{54 pixels}                   & \multicolumn{2}{l|}{$\sim$1900 pixels}              \\ \hline
\textbf{Separation Between Spectra} & \multicolumn{2}{l|}{6 pixels}                    & \multicolumn{2}{l|}{5-7 pixels}                     \\ \hline
\end{tabular}
\end{center}
\end{table}

\newpage
\section{OPTICAL DESIGN OVERVIEW}

The SCALES instrument is optimized for exoplanet imaging and must do the following:
\begin{itemize}
    \item Use Keck’s best adaptive optics configuration for high-contrast imaging
    \item Preserve the excellent image quality produced by Keck adaptive optics
    \item Use coronagraphy to suppress starlight
    \item Provide maximum sensitivity by following best practices for IR instrumentation
    \item Operate in the desired wavelength range and with the appropriate spectral resolutions for exoplanet characterization
    \item Without compromising the exoplanet imaging goals, the instrument should provide a versatile system for 1-5$\mu$m adaptive optics science.
\end{itemize}
These driving principles have led us to a top-level optical design that includes:
\begin{itemize}
    \item \textbf{Fore-optics}
        Two relays which, in order, make (1) a pupil plane used for a rotating cold-stop, (2) a focal plane used for insertable coronagraphic masks, (3) a pupil plane used for insertable Lyot stops, and (4) a focal plane sampled by a lenslet array for integral field spectroscopy. The optics are all-reflective, except for the coronagraph and lenslet array.  Immediately after the first pupil plane are a pair of filter wheels that are primarily used by the imager.
    \item \textbf{Imager}
        An insertable mirror at the Lyot stop plane diverts light away from the lenslet array and towards the imaging channel.  The imaging channel consists of an off-axis hyperboloid and a flat mirror, which image onto an H2RG detector.  
    \item \textbf{Spectrograph} 
        A collimator and camera that reimage the lenslet array spots onto an H2RG detector. 
        Between the collimator and camera are selectable reflective dispersers (prisms for low-resolution and gratings for medium resolution).  Between the camera and detector are selectable filters matched to each disperser.The optics are all reflective, except for the prisms and filters.
    \item \textbf{Slenslit slicer} 
        A set of optics can divert and re-insert light from the spectrograph into an image slicer (a `scenic bypass').  While the regular spectrograph is a typical lenslet-based low-resolution spectrograph, the combination of a lenslet + slicer (which we call a ``slenslit'') is a new type of integral field spectrograph that uses a lenslet array to sample the field (setting image quality in advance of the spectrograph), followed by a slicer that re-positions the 2D array of lenslet spots into a 1D pseudo-slit that can be dispersed over many more pixels than a standard lenslet-based IFS.  The optics are all reflective.
\end{itemize}

\begin{figure}[htp]
    \centering
    \includegraphics[width=0.95\textwidth]{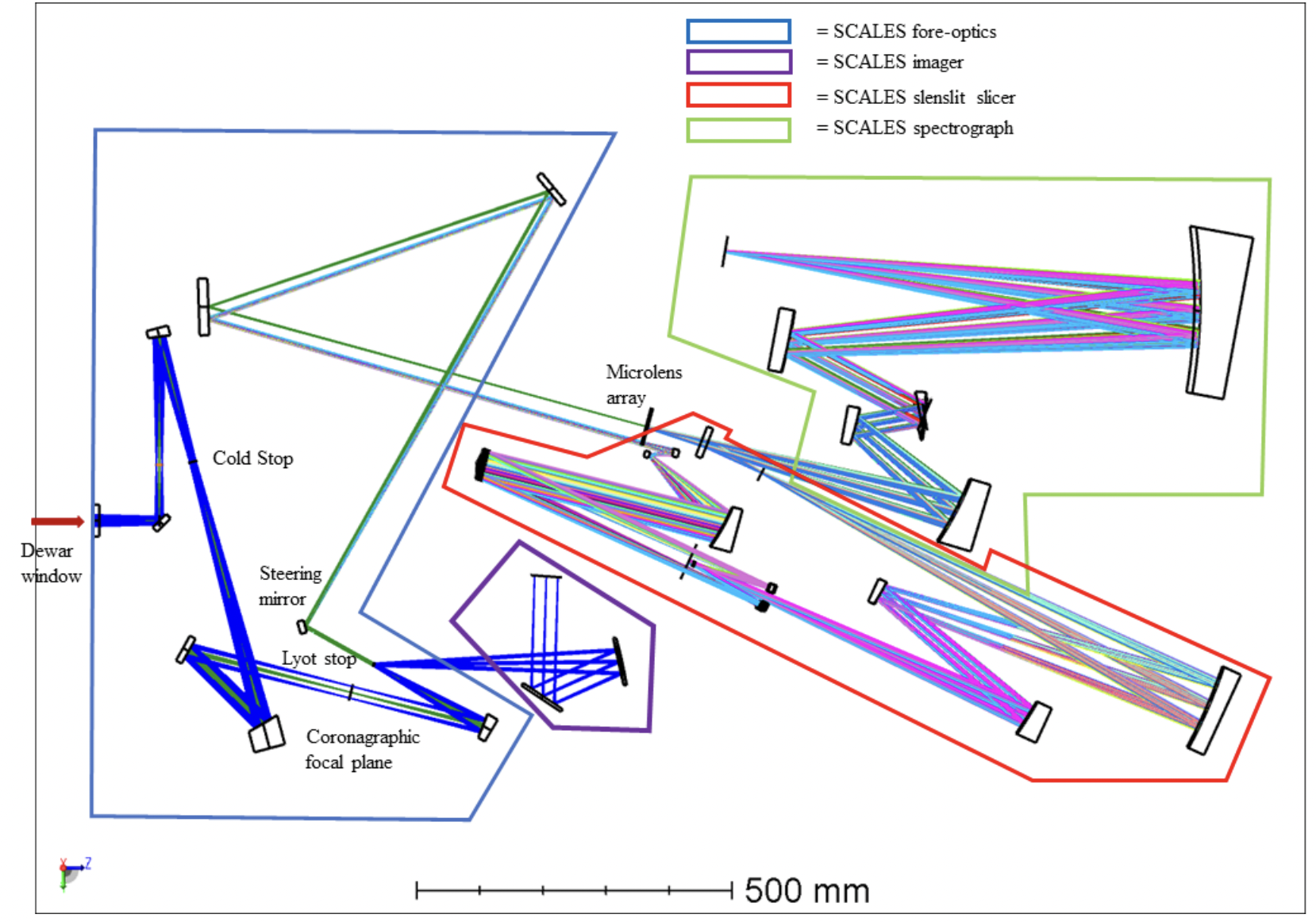}
    \caption[SCALES optical design.]{
        SCALES includes 4 major optical subsystems: a set of fore-optics feeding a lenslet array, an imaging channel, a low-resolution lenslet-based integral field spectrograph, and a medium-resolution ``slenslit'' spectrograph that sends the lenslet light through an image slicer and back into the main spectrograph as a pseudoslit of lenslet spots.
    }
    \label{figures:zemax-optical-design}
\end{figure}

A diagram showing the full optical layout, and denoting the above four subsystems, is shown in Figure~\ref{figures:zemax-optical-design}.  The SCALES optical design is described elsewhere in these proceedings \cite{Reni2022}, but notably, the design provides diffraction-limited performance at the lenslet array focal plane (21 nm maximum wavefront error), the imaging detector (56 nm maximum wavefront error) and the IFS detector (83 nm maximum wavefront error).  The throughput of the imager ranges from $\sim$55\%-70\%.  The throughput of the low-res IFS ranges from $\sim$50\%-60\%.  The throughput of the med-res IFS ranges from $\sim$35\%-45\%.

\section{INSTRUMENT LAYOUT}

\begin{figure}[h!]
    \centering
    \includegraphics[width=1.0\textwidth]{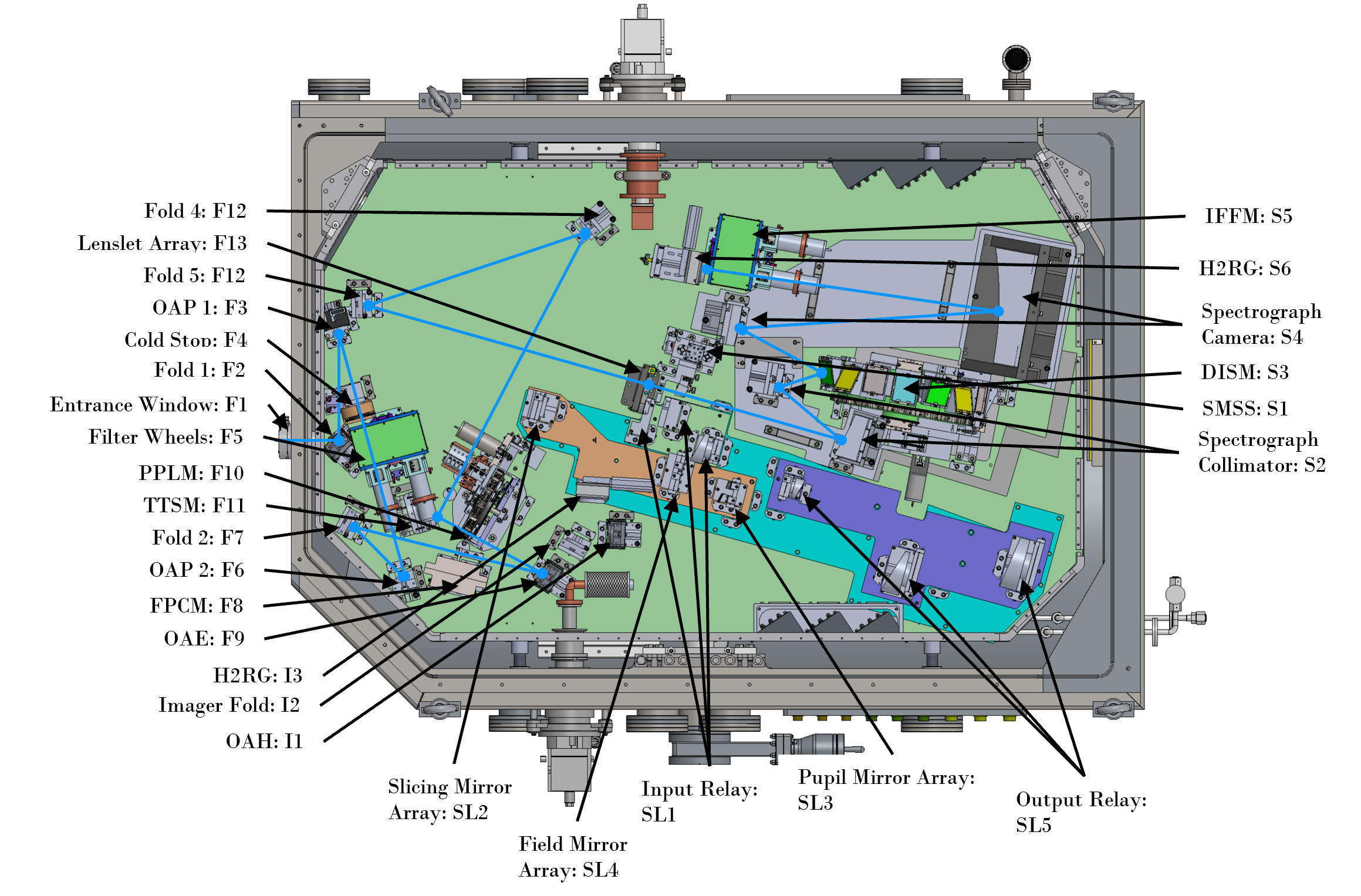}
    \caption[CAD rendering of the SCALES design.]{
        CAD rendering of the SCALES optical bench mounted in a cryostat. 
        An on-axis ray (highlighted in blue) traces the light through the low spectral resolution path, and optics are enumerated in the order that light travels as described in the text.

    }
    \label{figures:swx-opto-mechanical-design}
\end{figure}

\begin{figure}[h!]
    \centering
    \includegraphics[width=0.985\textwidth]{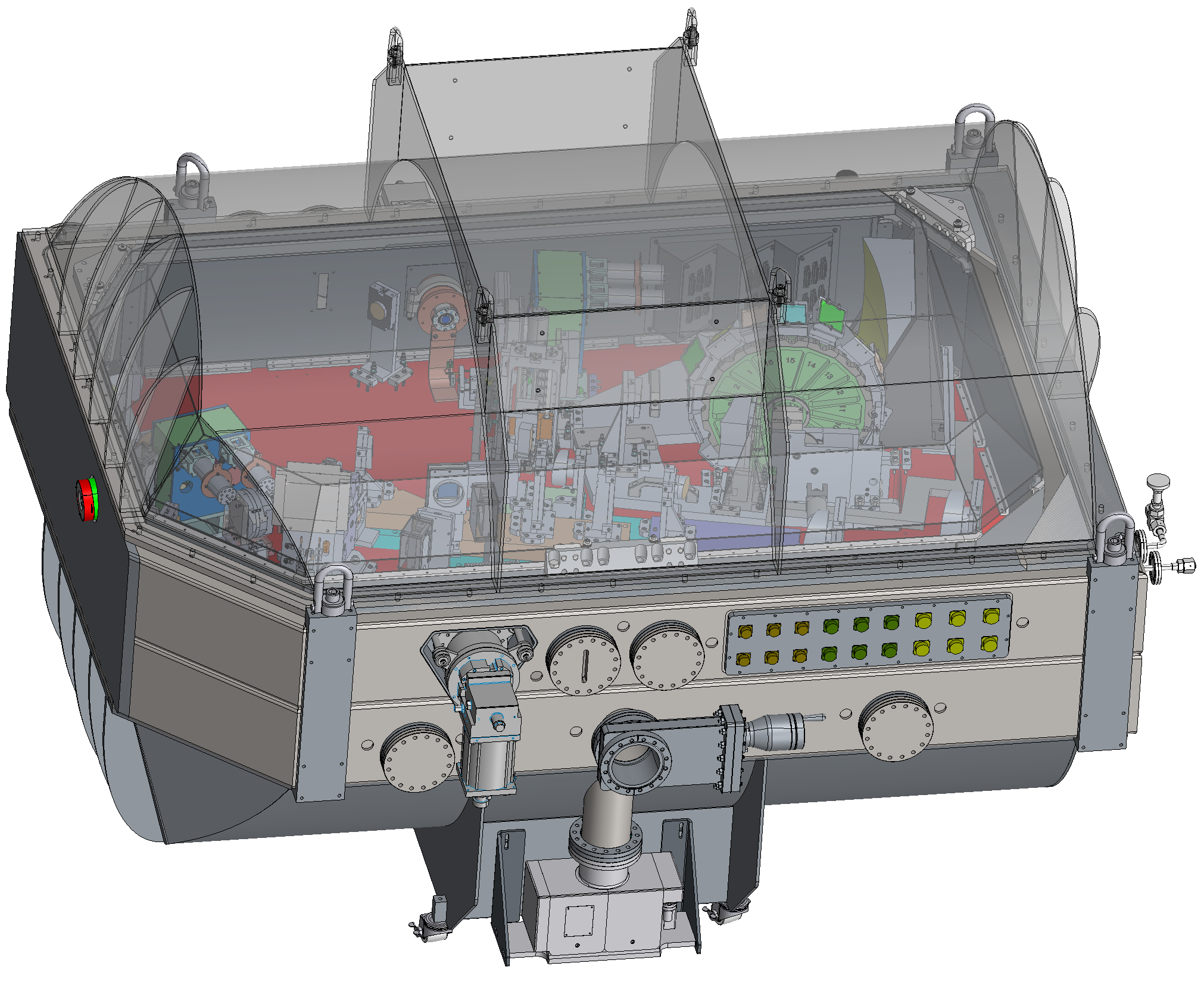}
    \caption{
SCALES cryostat with lid and heat shield lid made transparent.
    }
    \label{figures:cryostat}
\end{figure}

All of the SCALES optics are mounted on a single optics bench in a cryostat that is installed behind the Keck adaptive optics system.  In Figure~\ref{figures:swx-opto-mechanical-design}, we present a CAD rendering of the optics bench, and describe its different elements below.  In Figure \ref{figures:cryostat}, we present a CAD rendering of the full instrument.

Light path (mechanisms are bolded):

    \begin{center}
\textbf{Fore-Optics}
    \end{center}   
    \begin{itemize}
        \item F1) Cryostat entrance window: AR-coated CaF$_2$
       \item  F2) Fold-flat 1: This optic and all other reflective optics are diamond-turned RSA aluminum on an RSA aluminum mount
        \item F3) Off-axis-parabola 1: Collimates and makes a pupil plane
        \item  F4) \textbf{Cold-stop}: Rotating optic at a pupil plane that baffles light coming from outside of the Keck pupil.  The optic rotates to keep a fixed position angle, although it is generally left fixed for exoplanet imaging (Marois et al. 2007)
        \item F5) \textbf{Filter Wheels}: 2 wheels containing filters that are used with the imaging channel.
        \item F6) Off-axis-parabola 2: Focuses light and makes a focal plane
        \item F7) Fold 2
        \item F8) \textbf{Coronagraph Stage}: Linear stage at a focal plane holding selectable coronagraphs and an open position.
        \item F9) Off-axis-ellipse: Creates a pupil plane and a focal plane at the desired magnification for the lenslet array
        \item F10) \textbf{Lyot Stop Wheel}: Rotary mechanism at a pupil plane that holds selectable Lyot stops, neutral density filters, an open position, and a mirror to divert light to the imaging channel.
        \item F11) \textbf{Flat 3/tip-tilt Stage}: Precision steering mirror for placing and holding exoplanets and other objects at a desired location on the lenslet array.
        \item F12) Folds 4/5
        \item F13) Lenslet Array: Silicon lenslet array mated to a pinhole grid for diffraction suppression.  
        The lenslet array has two regions: one for the low-resolution IFS and one for the medium-resolution IFS.  
        On the input side, the lenslets are at a focal plane.  
        The pupil images made by the lenslet array become the focal plane for the spectrograph.
    \end{itemize}
    \begin{center}
\vspace{2cm}
\textbf{Imager}
    \end{center}    \begin{itemize}
        \item I1) Off-axis-hyperboloid: Creates a telecentric image at the detector focal plane at the appropriate magnification.
        \item I2) Imager Fold
        \item I3) Detector: $0.6-5.3 \mu$m-sensitive H2RG detector

    \end{itemize}
    \begin{center}
\textbf{Spectrograph}
    \end{center}
    \begin{itemize}
        \item  S1) \textbf{Mode Selector:} Linear stage that either lets the low-resolution lenslet light through to the spectrograph, or diverts and re-inserts the medium-resolution light to and from the slicer
        \item S2) Collimator 1/2: Two-mirror system to collimate the lenslet array light and make a pupil plane
        \item S3) \textbf{Disperser Carousel}: Rotary mechanism at a pupil plane with reflective dispersers on its outer edge.  
        The dispersers are LiF prisms for the low-resolution mode and gratings for the medium-resolution mode.
        \item S4) Camera 1/2: Two-mirror system to focus the collimated light to a focal plane
        \item S5) \textbf{IFS Filter Wheels:}
        Two Rotary mechanisms that contains bandpass filters for keeping the IFS spectra from overlapping, as well as blank and open positions
        \item S6) Detector: 0.6-5.3$\mu$m sensitive H2RG detector
    \end{itemize}
\begin{center}
\textbf{Slicer Optics}
\end{center}
    \begin{itemize}
        \item SL1) Input-relay: 3-element relay to make a focal plane at the image slicer
        \item SL2) Image slicer: divides the lenslet spots into rows
        \item SL3) Pupil mirrors: re-images the slices onto the field mirrors
        \item SL4) Field mirrors: steers the lenslet spots into a staggered pseudoslit
        \item SL5) Output-relay: 3-element relay to insert the slicer pseudoslit back into the spectrograph
    \end{itemize}

\section{PROJECT STATUS}
The SCALES project passed its preliminary design review for the baseline instrument in November 2021.  The imaging channel, which was considered an upgrade, held a separate preliminary design review in June 2022.  Subsequent final design reviews are on a subsystem-by-subsystem basis.  Some components have been built or acquired early as a risk reduction (described below).  The bulk of the purchasing is expected to begin in Fall 2022, with Integration and Testing beginning in late 2023 and shipping in 2025.  

\subsection{INTEGRATION AND TESTING FACILITIES}
SCALES will be integrated in a clean room at the UC Observatories Instrument shop in Santa Cruz.  The clean room will hold the SCALES cryostat when it arrives.  In the mean time, it is being set up to hold a liquid nitrogen cryostat for testing cryogenic mechanisms (Section \ref{mechanism testing}), a liquid nitrogen cryostat for testing optics (Section \ref{optic testing}) and a closed-cycle cryostat for testing detectors (Section \ref{detector testing}).  Additional cryogenic testing facilities are available at UC Irvine, UCLA, and the Indian Institute of Astrophysics.  

\subsection{EARLY DEVELOPMENT OF A CRYOGENIC MECHANISM}\label{mechanism testing}
\begin{figure}[h]
    \centering
    \includegraphics[width=0.4\textwidth]{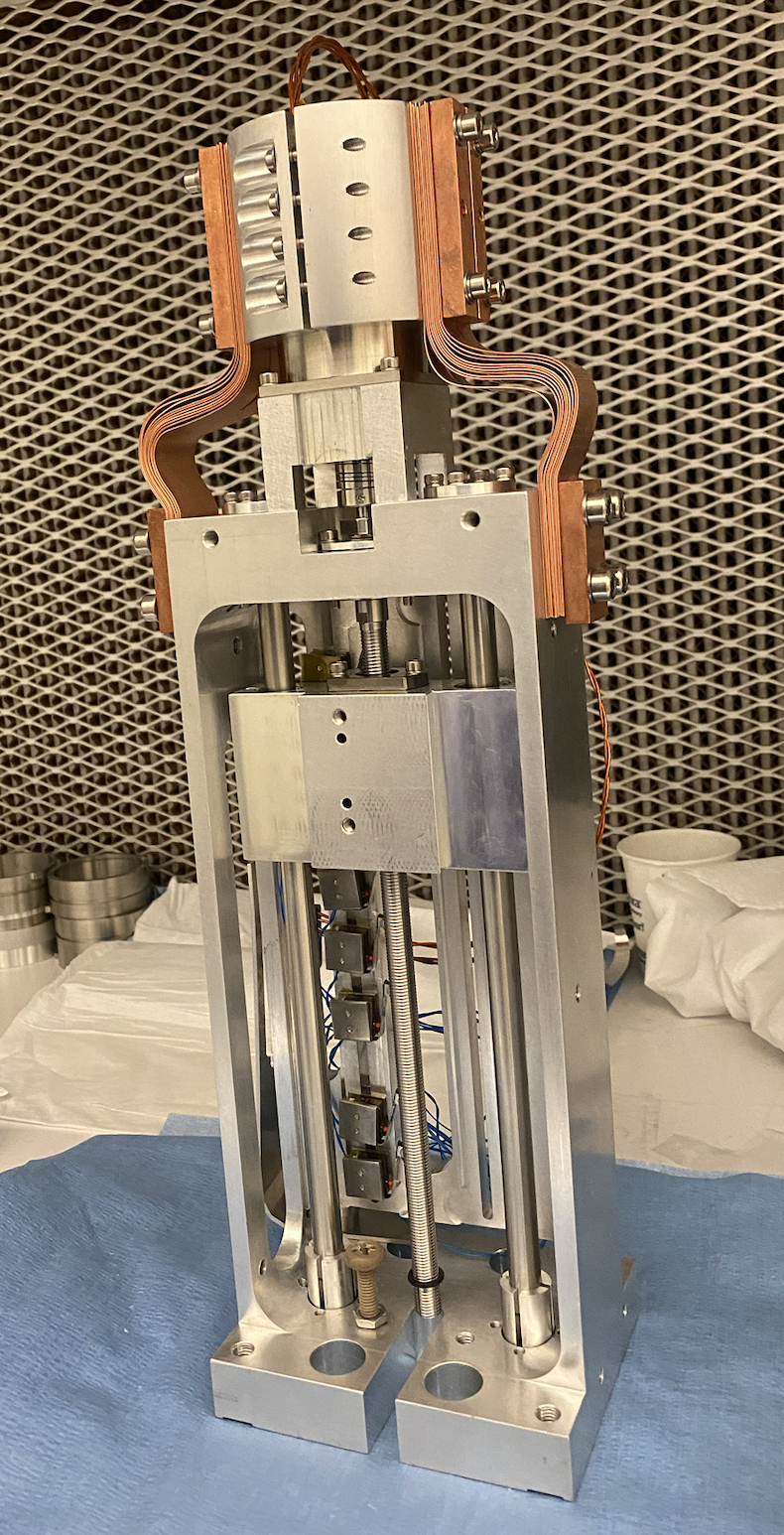}
        \caption{
        The completed coronagraph slide for SCALES.  The slide allows the selection and insertion of multiple coronagraphs, and has been successfully tested warm and cold.
    }
    \label{fig:coronagraph slide}
\end{figure}

SCALES has 10 cryogenic mechanisms, each of which involves a broad team of engineers (optical, mechanical, electrical, software) as they progress through design, fabrication, testing, and operations.  For SCALES, we made the decision to build one full cryogenic mechanism, the corongraph slide, early in the project, which allows our full team to iterate on the design.  Initial testing of the coronagraph slide was discussed in a previous SPIE Proceedings \cite{2021SPIE11823E..1WG}.  Since then, we have added a heat-sink to the motor, and are testing new Hall effect sensors to replace mechanical switches.  A photograph of the coronagraph slide is shown in Figure \ref{fig:coronagraph slide}.

\subsection{EARLY TESTING OF A DIAMOND-TURNED OPTIC}\label{optic testing}
With the exception of the entrance window, filters, dispersers, and the lenslet array, the SCALES optics are all diamond-turned aluminum.  Requirements for wavefront error and surface roughness are quite stringent, although multiple companies have the capabilities to meet our requirements.

\begin{figure}[h]
    \centering
    \includegraphics[width=0.985\textwidth]{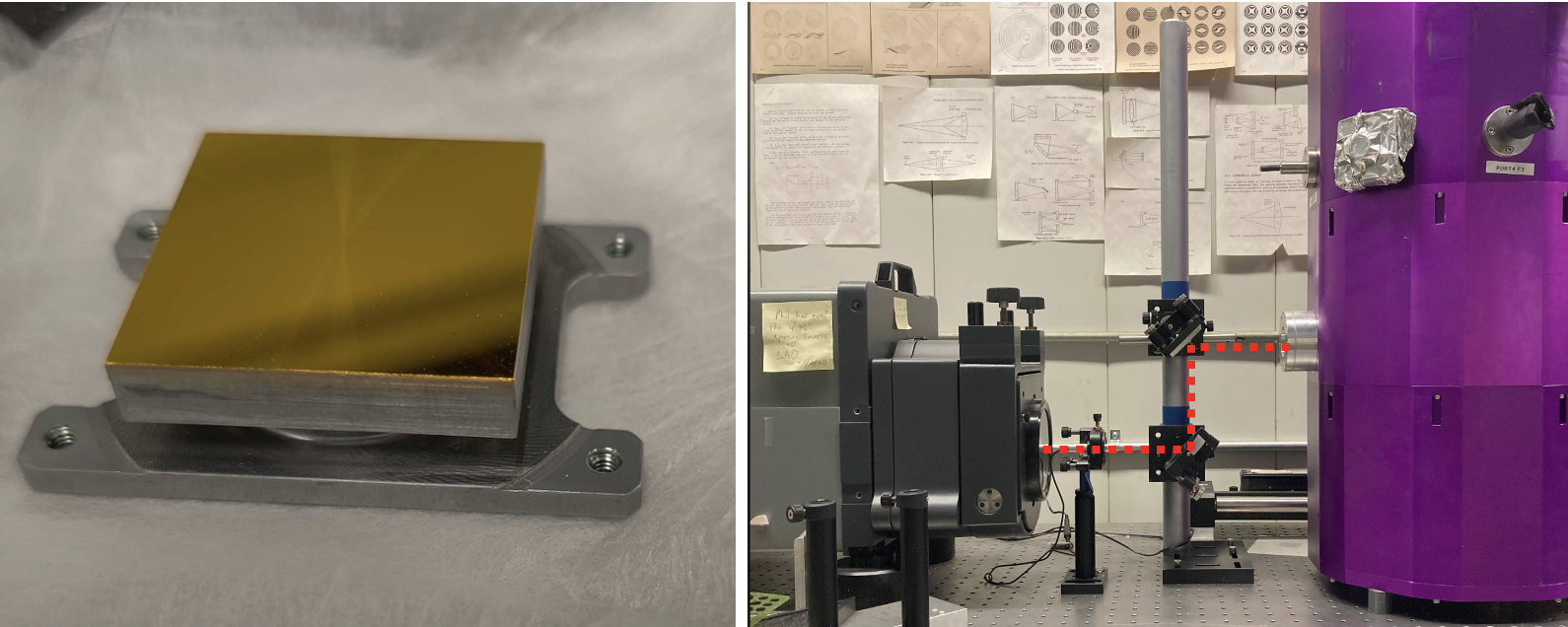}
    \caption{
   Left: A diamond-turned flat mirror that will be installed in SCALES.  Right: The cryogenic test setup for small diamond turned mirrors.  A Zygo interferometer sends light through a perisocope into a liquid nitrogen dewar that contains the diamond-turned flat mirror.  The test setup allows testing of the mirror at room temperature and operating temperature to check that the mirror does not distort when cooled.  
    }
    \label{fig:Son-X}
\end{figure}

The SCALES team is developing test setups for the small diamond-turned mirrors.  Among these measurements, we are measuring the wavefront error of each optic with a Zygo interferometer at room temperature and at operating temperature.  The test setup for measuring the optics at operating temperature is shown in Figure \ref{fig:Son-X}.

\subsection{EARLY DEVELOPMENT OF DETECTOR SUBSYSTEMS}\label{detector testing}
SCALES requires two infrared detectors: one for the integral field spectrograph and one for the imager.  The state of the art at SCALES' wavelengths are Teledyne H2RG detectors.  In July 2021, we received two detectors from the \textit{James Webb Space Telescope} project--SCA 17168 and SCA 17195, which were originally developed and characterized by the Goddard NIRSpec team\cite{2014PASP..126..739R}.  

Teledyne’s ground-based H2RGs feature 32 readout channels and the flexibility to run slow mode (typically 100 kpix/sec/output) or fast-mode (typically 2 Mpix/sec/output).  
This allows full-frame readouts as fast as 0.065 seconds.  
For JWST NIRSpec, the cable that connects the detector wire-bonds to a pinout is hardwired for 4 readout channels and slow mode.  
The minimum readout for JWST NIRSpec is 10.5 seconds.  
For ground-based infrared astronomy, the night sky is orders-of-magnitude brighter than it is in space and so faster readouts are generally required to avoid saturating on the sky background.

The SCALES team is contracting with Astroblank Scientific to speed up the detector readouts by using a hybrid buffered mode, where Teledyne's SIDECAR electronics use the fast-mode A2Ds to record slow-mode frames from the detector.  We expect to be able to read full frames in 1 second or less, which is sufficient to not saturate on sky background for the imager and integral field spectrograph.  Further development of subframes, to allow imaging of bright sources, is expected at a later date.


\acknowledgments 
 
We are grateful to the Heising-Simons Foundation, the Alfred P. Sloan Foundation, and the Mt. Cuba Astronomical Foundation for their generous support of our efforts.  This project also benefited from work conducted under NSF Grant 1608834 and the NSF Graduate Research Fellowship Program.  In addition, we thank the Robinson family and other private supporters, without whom this work would not be possible.  This work benefited from the 2022 Exoplanet Summer Program in the Other Worlds Laboratory (OWL) at the University of California, Santa Cruz, a program funded by the Heising-Simons Foundation.

\bibliography{main} 

\begin{thebibliography}{10}

\bibitem{2022arXiv220505696C}
{Currie}, T., {Biller}, B., {Lagrange}, A.-M., {Marois}, C., {Guyon}, O.,
  {Nielsen}, E., {Bonnefoy}, M., and {De Rosa}, R., ``{Direct Imaging and
  Spectroscopy of Extrasolar Planets},'' {\em arXiv e-prints} ,
  arXiv:2205.05696 (May 2022).

\bibitem{2014PNAS..11112661M}
{Macintosh}, B., {Graham}, J.~R., {Ingraham}, P., {Konopacky}, Q., {Marois},
  C., {Perrin}, M., {Poyneer}, L., {Bauman}, B., {Barman}, T., {Burrows},
  A.~S., {Cardwell}, A., {Chilcote}, J., {De Rosa}, R.~J., {Dillon}, D.,
  {Doyon}, R., {Dunn}, J., {Erikson}, D., {Fitzgerald}, M.~P., {Gavel}, D.,
  {Goodsell}, S., {Hartung}, M., {Hibon}, P., {Kalas}, P., {Larkin}, J.,
  {Maire}, J., {Marchis}, F., {Marley}, M.~S., {McBride}, J.,
  {Millar-Blanchaer}, M., {Morzinski}, K., {Norton}, A., {Oppenheimer}, B.~R.,
  {Palmer}, D., {Patience}, J., {Pueyo}, L., {Rantakyro}, F., {Sadakuni}, N.,
  {Saddlemyer}, L., {Savransky}, D., {Serio}, A., {Soummer}, R.,
  {Sivaramakrishnan}, A., {Song}, I., {Thomas}, S., {Wallace}, J.~K.,
  {Wiktorowicz}, S., and {Wolff}, S., ``{First light of the Gemini Planet
  Imager},'' {\em Proceedings of the National Academy of Science}~{\bf 111},
  12661--12666 (Sept. 2014).

\bibitem{2008SPIE.7014E..18B}
{Beuzit}, J.-L., {Feldt}, M., {Dohlen}, K., {Mouillet}, D., {Puget}, P.,
  {Wildi}, F., {Abe}, L., {Antichi}, J., {Baruffolo}, A., {Baudoz}, P.,
  {Boccaletti}, A., {Carbillet}, M., {Charton}, J., {Claudi}, R., {Downing},
  M., {Fabron}, C., {Feautrier}, P., {Fedrigo}, E., {Fusco}, T., {Gach}, J.-L.,
  {Gratton}, R., {Henning}, T., {Hubin}, N., {Joos}, F., {Kasper}, M.,
  {Langlois}, M., {Lenzen}, R., {Moutou}, C., {Pavlov}, A., {Petit}, C.,
  {Pragt}, J., {Rabou}, P., {Rigal}, F., {Roelfsema}, R., {Rousset}, G.,
  {Saisse}, M., {Schmid}, H.-M., {Stadler}, E., {Thalmann}, C., {Turatto}, M.,
  {Udry}, S., {Vakili}, F., and {Waters}, R., ``{SPHERE: a 'Planet Finder'
  instrument for the VLT},'' in [{\em Ground-based and Airborne Instrumentation
  for Astronomy II}{\nolinebreak\hspace{0.1em}]},  {McLean}, I.~S. and
  {Casali}, M.~M., eds., {\em Society of Photo-Optical Instrumentation
  Engineers (SPIE) Conference Series} {\bf 7014},  701418 (July 2008).

\bibitem{2015SPIE.9605E..1CG}
{Groff}, T.~D., {Kasdin}, N.~J., {Limbach}, M.~A., {Galvin}, M., {Carr}, M.~A.,
  {Knapp}, G., {Brandt}, T., {Loomis}, C., {Jarosik}, N., {Mede}, K.,
  {McElwain}, M.~W., {Leviton}, D.~B., {Miller}, K.~H., {Quijada}, M.~A.,
  {Guyon}, O., {Jovanovic}, N., {Takato}, N., and {Hayashi}, M., ``{The CHARIS
  IFS for high contrast imaging at Subaru},'' in [{\em Techniques and
  Instrumentation for Detection of Exoplanets
  VII}{\nolinebreak\hspace{0.1em}]},  {Shaklan}, S., ed., {\em Society of
  Photo-Optical Instrumentation Engineers (SPIE) Conference Series} {\bf 9605},
   96051C (Sept. 2015).

\bibitem{2018SPIE10702E..A5S}
{Skemer}, A.~J., {Stelter}, D., {Mawet}, D., {Fitzgerald}, M., {Mazin}, B.,
  {Guyon}, O., {Marois}, C., {Briesemeister}, Z., {Brandt}, T., {Chilcote}, J.,
  {Delorme}, J.-R., {Jovanovic}, N., {Lu}, J., {Millar-Blanchaer}, M.,
  {Wallace}, J., {Vasisht}, G., {Roberts}, L.~C., and {Wang}, J., ``{The
  planetary systems imager: 2-5 micron channel},'' in [{\em Ground-based and
  Airborne Instrumentation for Astronomy VII}{\nolinebreak\hspace{0.1em}]},
  {Evans}, C.~J., {Simard}, L., and {Takami}, H., eds., {\em Society of
  Photo-Optical Instrumentation Engineers (SPIE) Conference Series} {\bf
  10702},  10702A5 (July 2018).

\bibitem{2020SPIE11447E..64S}
{Stelter}, R.~D., {Skemer}, A.~J., {Sallum}, S., {Kupke}, R., {Hinz}, P.,
  {Mawet}, D., {Jensen-Clem}, R., {Ratliffe}, C., {MacDonald}, N., {Deich}, W.,
  {Kruglikov}, G., {Kassis}, M., {Lyke}, J., {Briesemeister}, Z., {Miles}, B.,
  {Gerard}, B., {Fitzgerald}, M., {Brandt}, T., and {Marois}, C., ``{Update on
  the preliminary design of SCALES: the Santa Cruz Array of Lenslets for
  Exoplanet Spectroscopy},'' in [{\em Society of Photo-Optical Instrumentation
  Engineers (SPIE) Conference Series}{\nolinebreak\hspace{0.1em}]},  {\em
  Society of Photo-Optical Instrumentation Engineers (SPIE) Conference Series}
  {\bf 11447},  1144764 (Dec. 2020).

\bibitem{2015SPIE.9605E..1DS}
{Skemer}, A.~J., {Hinz}, P., {Montoya}, M., {Skrutskie}, M.~F., {Leisenring},
  J., {Durney}, O., {Woodward}, C.~E., {Wilson}, J., {Nelson}, M., {Bailey},
  V., {Defrere}, D., and {Stone}, J., ``{First light with ALES: A 2-5 micron
  adaptive optics Integral Field Spectrograph for the LBT},'' in [{\em
  Techniques and Instrumentation for Detection of Exoplanets
  VII}{\nolinebreak\hspace{0.1em}]},  {Shaklan}, S., ed., {\em Society of
  Photo-Optical Instrumentation Engineers (SPIE) Conference Series} {\bf 9605},
   96051D (Sept. 2015).

\bibitem{2018SPIE10702E..3FS}
{Stone}, J.~M., {Skemer}, A.~J., {Hinz}, P., {Briesemeister}, Z., {Barman}, T.,
  {Woodward}, C.~E., {Skrutskie}, M., and {Leisenring}, J., ``{On-sky
  operations with the ALES integral field spectrograph},'' in [{\em
  Ground-based and Airborne Instrumentation for Astronomy
  VII}{\nolinebreak\hspace{0.1em}]},  {Evans}, C.~J., {Simard}, L., and
  {Takami}, H., eds., {\em Society of Photo-Optical Instrumentation Engineers
  (SPIE) Conference Series} {\bf 10702},  107023F (July 2018).

\bibitem{Reni2022}
Kupke, R., Stelter, R.~D., Hasan, A., Surya, A., Kain, I., Briesemeister, Z.,
  Li, J., Hinz, P., Skemer, A., Gerard, B., Dillon, D., and Ratliff, C.,
  ``Scales for keck: Optical design,'' in [{\em Ground-based and Airborne
  Instrumentation for Astronomy IX}{\nolinebreak\hspace{0.1em}]},   {\bf
  12184}, SPIE (2022).

\bibitem{Banyal2022}
Banyal, R.~K., Hasan, A., Kupke, R., Sivarani, T., Skemer, A.~J., MacDonald,
  N., Sallum, S., Deich, W., Divakar, D.~K., Fitzgerald, M., Govinda, K.~V.,
  Prakaesh, A., Ratliff, C., Sethuram, R., Sriram, S., Stelter, D., Surya, A.,
  Varshney, H.~M., and Wang, E., ``Design of an ir imaging channel for the keck
  observatory scales instrument,'' in [{\em Advances in Optical and Mechanical
  Technologies for Telescopes and Instrumentation
  V}{\nolinebreak\hspace{0.1em}]},   {\bf 12188}, SPIE (2022).

\bibitem{DenoSlenslit2022}
Stelter, D., Skemer, A., Kupke, R., and Bourgenot, C., ``Weighing
  exo-atmospheres: A novel mid-resolution spectral mode for scales,'' in [{\em
  Ground-based and Airborne Instrumentation for Astronomy
  IX}{\nolinebreak\hspace{0.1em}]},   {\bf 12184}, SPIE (2022).

\bibitem{Jialin2022}
Li, J., Skemer, A., van Kooten, M., Kupke, R., and MacDonald, N., ``Fabrication
  of pupil masks for a new infrared exoplanet imager at keck observatory,'' in
  [{\em Adaptive Optics Systems VIII}{\nolinebreak\hspace{0.1em}]},   {\bf
  12185}, SPIE (2022).

\bibitem{Lach2022}
Lach, M., Sallum, S., and Skemer, A., ``Simulating the performance of aperture
  mask designs for scales,'' in [{\em Ground-based and Airborne Instrumentation
  for Astronomy IX}{\nolinebreak\hspace{0.1em}]},   {\bf 12183}, SPIE (2022).

\bibitem{Kassis2022}
Kassis, M., Allen, S.~L., Alverez, C., Baker, A., Banyal, R.~K., Bertz, R.,
  Beichman, C., Brown, A., Brown, M., Bundy, K., Cabak, G., Cetre, S., Chin,
  J., Chun, M.~R., Cooke, J., Delorme, J., Deich, W., Dekany, R.~G., Devenot,
  M., Doppmann, G., Edelstein, J., Fitzgerald, M.~P., Fucik, J.~R., Gao, M.,
  Gibson, S., Gillingham, P.~R., Gomez, P., Gottschalk, C., Sam~Halverson,
  G.~H., Hinz, P., Holden, B.~P., Howard, A.~W., Jones, T., Jovanovic, N.,
  Kirby, E., Krishnan, S., Kupke, R., Lanclos, K., Larkin, J.~E., Leifer,
  S.~D., Lewis, H.~A., Lilley, S., Lu, J.~R., Lyke, J.~E., MacDonald, N.,
  Martin, C., Mather, J., Matuszewski, M., Mawet, D., McCarney, B., McGurk, R.,
  Marin, E., Millar-Blanchaer, M.~A., Nance, C., Nash, R.~B., Neill, J.~D.,
  O’Meara, J.~M., Peretz, E., Poppett, C., Konopacky, Q., Radovan, M.~V.,
  Ragland, S., Rider, K., Roberts, M., Rockosi, C., Roy, A., Rubenzahl, R.,
  Sallum, S., Sandford, D., Savage, M., Shen, B., Simha, S., Skemer, A.~J.,
  Steidel, C.~C., Stelter, D., Surendran, A., Thorne, J., Walawender, J.,
  Westfall, K.~B., Wizinowich, P., Vahala, K., Wright, S., Wold, T., and Yeh,
  S., ``Innovations and advances in instrumentation at the w. m. keck
  observatory, vol. ii,'' in [{\em Ground-based and Airborne Instrumentation
  for Astronomy IX}{\nolinebreak\hspace{0.1em}]},   {\bf 12184}, SPIE (2022).

\bibitem{2014ApJ...787...78M}
{Morley}, C.~V., {Marley}, M.~S., {Fortney}, J.~J., {Lupu}, R., {Saumon}, D.,
  {Greene}, T., and {Lodders}, K., ``{Water Clouds in Y Dwarfs and
  Exoplanets},'' {\em ApJ}~{\bf 787},  78 (May 2014).

\bibitem{2014SPIE.9148E..0UM}
{Marois}, C., {Correia}, C., {Galicher}, R., {Ingraham}, P., {Macintosh}, B.,
  {Currie}, T., and {De Rosa}, R., ``{GPI PSF subtraction with TLOCI: the next
  evolution in exoplanet/disk high-contrast imaging},'' in [{\em Adaptive
  Optics Systems IV}{\nolinebreak\hspace{0.1em}]},  {Marchetti}, E., {Close},
  L.~M., and {Vran}, J.-P., eds., {\em Society of Photo-Optical Instrumentation
  Engineers (SPIE) Conference Series} {\bf 9148},  91480U (July 2014).

\bibitem{2015MNRAS.454..129V}
{Vigan}, A., {Gry}, C., {Salter}, G., {Mesa}, D., {Homeier}, D., {Moutou}, C.,
  and {Allard}, F., ``{High-contrast imaging of Sirius A with VLT/SPHERE:
  looking for giant planets down to one astronomical unit},'' {\em MNRAS}~{\bf
  454},  129--143 (Nov. 2015).

\bibitem{SPHERE_manual}
{Wahhaj}, Z., {Milli}, J., {Rodler}, F., {Gerard}, J., {Vigan}, A., {van den
  Ancker}, M., and {Boffin}, H., ``{Very Large Telescope SPHERE User Manual},''
  (Sept. 2019).

\bibitem{2014ApJ...792...17S}
{Skemer}, A.~J., {Marley}, M.~S., {Hinz}, P.~M., {Morzinski}, K.~M.,
  {Skrutskie}, M.~F., {Leisenring}, J.~M., {Close}, L.~M., {Saumon}, D.,
  {Bailey}, V.~P., {Briguglio}, R., {Defrere}, D., {Esposito}, S., {Follette},
  K.~B., {Hill}, J.~M., {Males}, J.~R., {Puglisi}, A., {Rodigas}, T.~J., and
  {Xompero}, M., ``{Directly Imaged L-T Transition Exoplanets in the
  Mid-infrared},'' {\em ApJ}~{\bf 792},  17 (Sept. 2014).

\bibitem{2020SPIE11447E..4ZB}
{Briesemeister}, Z., {Sallum}, S., {Skemer}, A., {Stelter}, R.~D., {Hinz}, P.,
  and {Brandt}, T., ``{End-to-end simulation of the SCALES integral field
  spectrograph},'' in [{\em Society of Photo-Optical Instrumentation Engineers
  (SPIE) Conference Series}{\nolinebreak\hspace{0.1em}]},  {\em Society of
  Photo-Optical Instrumentation Engineers (SPIE) Conference Series} {\bf
  11447},  114474Z (Dec. 2020).

\bibitem{2018SPIE10702E..2QB}
{Briesemeister}, Z., {Skemer}, A.~J., {Stone}, J.~M., {Stelter}, R.~D., {Hinz},
  P., {Leisenring}, J., {Skrutskie}, M.~F., {Woodward}, C.~E., and {Barman},
  T., ``{MEAD: data reduction pipeline for ALES integral field spectrograph and
  LBTI thermal infrared calibration unit},'' in [{\em Ground-based and Airborne
  Instrumentation for Astronomy VII}{\nolinebreak\hspace{0.1em}]},  {Evans},
  C.~J., {Simard}, L., and {Takami}, H., eds., {\em Society of Photo-Optical
  Instrumentation Engineers (SPIE) Conference Series} {\bf 10702},  107022Q
  (July 2018).

\bibitem{2014ApJ...797...14P}
{Perryman}, M., {Hartman}, J., {Bakos}, G.~{\'A}., and {Lindegren}, L.,
  ``{Astrometric Exoplanet Detection with Gaia},'' {\em ApJ}~{\bf 797},  14
  (Dec. 2014).

\bibitem{2019AJ....158..140B}
{Brandt}, T.~D., {Dupuy}, T.~J., and {Bowler}, B.~P., ``{Precise Dynamical
  Masses of Directly Imaged Companions from Relative Astrometry, Radial
  Velocities, and Hipparcos-Gaia DR2 Accelerations},'' {\em AJ}~{\bf 158},  140
  (Oct. 2019).

\bibitem{2016ApJ...826L..17S}
{Skemer}, A.~J., {Morley}, C.~V., {Allers}, K.~N., {Geballe}, T.~R., {Marley},
  M.~S., {Fortney}, J.~J., {Faherty}, J.~K., {Bjoraker}, G.~L., and {Lupu}, R.,
  ``{The First Spectrum of the Coldest Brown Dwarf},'' {\em ApJl}~{\bf 826},
  L17 (Aug. 2016).

\bibitem{2015Sci...350...64M}
{Macintosh}, B., {Graham}, J.~R., {Barman}, T., {De Rosa}, R.~J., {Konopacky},
  Q., {Marley}, M.~S., {Marois}, C., {Nielsen}, E.~L., {Pueyo}, L., {Rajan},
  A., {Rameau}, J., {Saumon}, D., {Wang}, J.~J., {Patience}, J., {Ammons}, M.,
  {Arriaga}, P., {Artigau}, E., {Beckwith}, S., {Brewster}, J., {Bruzzone}, S.,
  {Bulger}, J., {Burningham}, B., {Burrows}, A.~S., {Chen}, C., {Chiang}, E.,
  {Chilcote}, J.~K., {Dawson}, R.~I., {Dong}, R., {Doyon}, R., {Draper}, Z.~H.,
  {Duch{\^e}ne}, G., {Esposito}, T.~M., {Fabrycky}, D., {Fitzgerald}, M.~P.,
  {Follette}, K.~B., {Fortney}, J.~J., {Gerard}, B., {Goodsell}, S.,
  {Greenbaum}, A.~Z., {Hibon}, P., {Hinkley}, S., {Cotten}, T.~H., {Hung},
  L.~W., {Ingraham}, P., {Johnson-Groh}, M., {Kalas}, P., {Lafreniere}, D.,
  {Larkin}, J.~E., {Lee}, J., {Line}, M., {Long}, D., {Maire}, J., {Marchis},
  F., {Matthews}, B.~C., {Max}, C.~E., {Metchev}, S., {Millar-Blanchaer},
  M.~A., {Mittal}, T., {Morley}, C.~V., {Morzinski}, K.~M., {Murray-Clay}, R.,
  {Oppenheimer}, R., {Palmer}, D.~W., {Patel}, R., {Perrin}, M.~D., {Poyneer},
  L.~A., {Rafikov}, R.~R., {Rantakyr{\"o}}, F.~T., {Rice}, E.~L., {Rojo}, P.,
  {Rudy}, A.~R., {Ruffio}, J.~B., {Ruiz}, M.~T., {Sadakuni}, N., {Saddlemyer},
  L., {Salama}, M., {Savransky}, D., {Schneider}, A.~C., {Sivaramakrishnan},
  A., {Song}, I., {Soummer}, R., {Thomas}, S., {Vasisht}, G., {Wallace}, J.~K.,
  {Ward-Duong}, K., {Wiktorowicz}, S.~J., {Wolff}, S.~G., and {Zuckerman}, B.,
  ``{Discovery and spectroscopy of the young jovian planet 51 Eri b with the
  Gemini Planet Imager},'' {\em Science}~{\bf 350},  64--67 (Oct. 2015).

\bibitem{2019SPIE11117E..0WJ}
{Jensen-Clem}, R., {Bond}, C.~Z., {Cetre}, S., {McEwen}, E., {Wizinowich}, P.,
  {Ragland}, S., {Mawet}, D., and {Graham}, J., ``{Demonstrating predictive
  wavefront control with the Keck II near-infrared pyramid wavefront sensor},''
  in [{\em Society of Photo-Optical Instrumentation Engineers (SPIE) Conference
  Series}{\nolinebreak\hspace{0.1em}]},  {\em Society of Photo-Optical
  Instrumentation Engineers (SPIE) Conference Series} {\bf 11117},  111170W
  (Sept. 2019).

\bibitem{2022JATIS...8b9006V}
{van Kooten}, M. A.~M., {Jensen-Clem}, R., {Cetre}, S., {Ragland}, S., {Bond},
  C.~Z., {Fowler}, J., and {Wizinowich}, P., ``{Predictive wavefront control on
  Keck II adaptive optics bench: on-sky coronagraphic results},'' {\em Journal
  of Astronomical Telescopes, Instruments, and Systems}~{\bf 8},  029006 (Apr.
  2022).

\bibitem{Hinz2022}
et~al., P.~H., ``An asm-based ao system for w.m. keck observatory,'' in [{\em
  Ground-based and Airborne Instrumentation for Astronomy
  IX}{\nolinebreak\hspace{0.1em}]},   {\bf 12185}, SPIE (2022).

\bibitem{2012ApJ...745....5K}
{Kraus}, A.~L. and {Ireland}, M.~J., ``{LkCa 15: A Young Exoplanet Caught at
  Formation?},'' {\em ApJ}~{\bf 745},  5 (Jan. 2012).

\bibitem{2018A&A...617A..44K}
{Keppler}, M., {Benisty}, M., {M{\"u}ller}, A., {Henning}, T., {van Boekel},
  R., {Cantalloube}, F., {Ginski}, C., {van Holstein}, R.~G., {Maire}, A.~L.,
  {Pohl}, A., {Samland}, M., {Avenhaus}, H., {Baudino}, J.~L., {Boccaletti},
  A., {de Boer}, J., {Bonnefoy}, M., {Chauvin}, G., {Desidera}, S., {Langlois},
  M., {Lazzoni}, C., {Marleau}, G.~D., {Mordasini}, C., {Pawellek}, N.,
  {Stolker}, T., {Vigan}, A., {Zurlo}, A., {Birnstiel}, T., {Brandner}, W.,
  {Feldt}, M., {Flock}, M., {Girard}, J., {Gratton}, R., {Hagelberg}, J.,
  {Isella}, A., {Janson}, M., {Juhasz}, A., {Kemmer}, J., {Kral}, Q.,
  {Lagrange}, A.~M., {Launhardt}, R., {Matter}, A., {M{\'e}nard}, F., {Milli},
  J., {Molli{\`e}re}, P., {Olofsson}, J., {P{\'e}rez}, L., {Pinilla}, P.,
  {Pinte}, C., {Quanz}, S.~P., {Schmidt}, T., {Udry}, S., {Wahhaj}, Z.,
  {Williams}, J.~P., {Buenzli}, E., {Cudel}, M., {Dominik}, C., {Galicher}, R.,
  {Kasper}, M., {Lannier}, J., {Mesa}, D., {Mouillet}, D., {Peretti}, S.,
  {Perrot}, C., {Salter}, G., {Sissa}, E., {Wildi}, F., {Abe}, L., {Antichi},
  J., {Augereau}, J.~C., {Baruffolo}, A., {Baudoz}, P., {Bazzon}, A., {Beuzit},
  J.~L., {Blanchard}, P., {Brems}, S.~S., {Buey}, T., {De Caprio}, V.,
  {Carbillet}, M., {Carle}, M., {Cascone}, E., {Cheetham}, A., {Claudi}, R.,
  {Costille}, A., {Delboulb{\'e}}, A., {Dohlen}, K., {Fantinel}, D.,
  {Feautrier}, P., {Fusco}, T., {Giro}, E., {Gluck}, L., {Gry}, C., {Hubin},
  N., {Hugot}, E., {Jaquet}, M., {Le Mignant}, D., {Llored}, M., {Madec}, F.,
  {Magnard}, Y., {Martinez}, P., {Maurel}, D., {Meyer}, M.,
  {M{\"o}ller-Nilsson}, O., {Moulin}, T., {Mugnier}, L., {Orign{\'e}}, A.,
  {Pavlov}, A., {Perret}, D., {Petit}, C., {Pragt}, J., {Puget}, P., {Rabou},
  P., {Ramos}, J., {Rigal}, F., {Rochat}, S., {Roelfsema}, R., {Rousset}, G.,
  {Roux}, A., {Salasnich}, B., {Sauvage}, J.~F., {Sevin}, A., {Soenke}, C.,
  {Stadler}, E., {Suarez}, M., {Turatto}, M., and {Weber}, L., ``{Discovery of
  a planetary-mass companion within the gap of the transition disk around PDS
  70},'' {\em A\&A}~{\bf 617},  A44 (Sept. 2018).

\bibitem{2016SPIE.9907E..0DS}
{Sallum}, S., {Eisner}, J., {Close}, L.~M., {Hinz}, P.~M., {Follette}, K.~B.,
  {Kratter}, K., {Skemer}, A.~J., {Bailey}, V.~P., {Briguglio}, R., {Defrere},
  D., {Macintosh}, B.~A., {Males}, J.~R., {Morzinski}, K.~M., {Puglisi}, A.~T.,
  {Rodigas}, T.~J., {Spalding}, E., {Tuthill}, P.~G., {Vaz}, A., {Weinberger},
  A., and {Xomperio}, M., ``{Imaging protoplanets: observing transition disks
  with non-redundant masking},'' in [{\em Optical and Infrared Interferometry
  and Imaging V}{\nolinebreak\hspace{0.1em}]},  {Malbet}, F., {Creech-Eakman},
  M.~J., and {Tuthill}, P.~G., eds., {\em Society of Photo-Optical
  Instrumentation Engineers (SPIE) Conference Series} {\bf 9907},  99070D (Aug.
  2016).

\bibitem{2015Natur.527..342S}
{Sallum}, S., {Follette}, K.~B., {Eisner}, J.~A., {Close}, L.~M., {Hinz}, P.,
  {Kratter}, K., {Males}, J., {Skemer}, A., {Macintosh}, B., {Tuthill}, P.,
  {Bailey}, V., {Defr{\`e}re}, D., {Morzinski}, K., {Rodigas}, T., {Spalding},
  E., {Vaz}, A., and {Weinberger}, A.~J., ``{Accreting protoplanets in the LkCa
  15 transition disk},'' {\em Nature}~{\bf 527},  342--344 (Nov. 2015).

\bibitem{2018SPIE10703E..1ZB}
{Bond}, C.~Z., {Wizinowich}, P., {Chun}, M., {Mawet}, D., {Lilley}, S.,
  {Cetre}, S., {Jovanovic}, N., {Delorme}, J.-R., {Wetherell}, E., {Jacobson},
  S., {Lockhart}, C., {Warmbier}, E., {Wallace}, J.~K., {Hall}, D.~N.,
  {Goebel}, S., {Guyon}, O., {Plantet}, C., {Agapito}, G., {Giordano}, C.,
  {Esposito}, S., and {Femenia-Castella}, B., ``{Adaptive optics with an
  infrared pyramid wavefront sensor},'' in [{\em Adaptive Optics Systems
  VI}{\nolinebreak\hspace{0.1em}]},  {Close}, L.~M., {Schreiber}, L., and
  {Schmidt}, D., eds., {\em Society of Photo-Optical Instrumentation Engineers
  (SPIE) Conference Series} {\bf 10703},  107031Z (July 2018).

\bibitem{2019NatAs...3..749H}
{Haffert}, S.~Y., {Bohn}, A.~J., {de Boer}, J., {Snellen}, I.~A.~G.,
  {Brinchmann}, J., {Girard}, J.~H., {Keller}, C.~U., and {Bacon}, R., ``{Two
  accreting protoplanets around the young star PDS 70},'' {\em Nature
  Astronomy}~{\bf 3},  749--754 (June 2019).

\bibitem{2011ApJ...733...65B}
{Barman}, T.~S., {Macintosh}, B., {Konopacky}, Q.~M., and {Marois}, C.,
  ``{Clouds and Chemistry in the Atmosphere of Extrasolar Planet HR8799b},''
  {\em ApJ}~{\bf 733},  65 (May 2011).

\bibitem{2012ApJ...754..135M}
{Marley}, M.~S., {Saumon}, D., {Cushing}, M., {Ackerman}, A.~S., {Fortney},
  J.~J., and {Freedman}, R., ``{Masses, Radii, and Cloud Properties of the HR
  8799 Planets},'' {\em ApJ}~{\bf 754},  135 (Aug. 2012).

\bibitem{2016ApJ...817..166S}
{Skemer}, A.~J., {Morley}, C.~V., {Zimmerman}, N.~T., {Skrutskie}, M.~F.,
  {Leisenring}, J., {Buenzli}, E., {Bonnefoy}, M., {Bailey}, V., {Hinz}, P.,
  {Defr{\'e}re}, D., {Esposito}, S., {Apai}, D., {Biller}, B., {Brandner}, W.,
  {Close}, L., {Crepp}, J.~R., {De Rosa}, R.~J., {Desidera}, S., {Eisner}, J.,
  {Fortney}, J., {Freedman}, R., {Henning}, T., {Hofmann}, K.-H., {Kopytova},
  T., {Lupu}, R., {Maire}, A.-L., {Males}, J.~R., {Marley}, M., {Morzinski},
  K., {Oza}, A., {Patience}, J., {Rajan}, A., {Rieke}, G., {Schertl}, D.,
  {Schlieder}, J., {Stone}, J., {Su}, K., {Vaz}, A., {Visscher}, C.,
  {Ward-Duong}, K., {Weigelt}, G., and {Woodward}, C.~E., ``{The LEECH
  Exoplanet Imaging Survey: Characterization of the Coldest Directly Imaged
  Exoplanet, GJ 504 b, and Evidence for Superstellar Metallicity},'' {\em
  ApJ}~{\bf 817},  166 (Feb. 2016).

\bibitem{2013cctp.book..367M}
{Marley}, M.~S., {Ackerman}, A.~S., {Cuzzi}, J.~N., and {Kitzmann}, D.,
  ``{Clouds and Hazes in Exoplanet Atmospheres},'' in [{\em Comparative
  Climatology of Terrestrial Planets}{\nolinebreak\hspace{0.1em}]},
  {Mackwell}, S.~J., {Simon-Miller}, A.~A., {Harder}, J.~W., and {Bullock},
  M.~A., eds.,  367--392 (2013).

\bibitem{2016ApJ...820...78L}
{Line}, M.~R. and {Parmentier}, V., ``{The Influence of Nonuniform Cloud Cover
  on Transit Transmission Spectra},'' {\em ApJ}~{\bf 820},  78 (Mar. 2016).

\bibitem{2013Sci...339.1398K}
{Konopacky}, Q.~M., {Barman}, T.~S., {Macintosh}, B.~A., and {Marois}, C.,
  ``{Detection of Carbon Monoxide and Water Absorption Lines in an Exoplanet
  Atmosphere},'' {\em Science}~{\bf 339},  1398--1401 (Mar. 2013).

\bibitem{2015ApJ...804...61B}
{Barman}, T.~S., {Konopacky}, Q.~M., {Macintosh}, B., and {Marois}, C.,
  ``{Simultaneous Detection of Water, Methane, and Carbon Monoxide in the
  Atmosphere of Exoplanet HR8799b},'' {\em ApJ}~{\bf 804},  61 (May 2015).

\bibitem{1995A&AS..113..347B}
{Bacon}, R., {Adam}, G., {Baranne}, A., {Courtes}, G., {Dubet}, D., {Dubois},
  J.~P., {Emsellem}, E., {Ferruit}, P., {Georgelin}, Y., {Monnet}, G.,
  {Pecontal}, E., {Rousset}, A., and {Say}, F., ``{3D spectrography at high
  spatial resolution. I. Concept and realization of the integral field
  spectrograph TIGER.},'' {\em A\&A Suppl.}~{\bf 113},  347 (Oct. 1995).

\bibitem{2021SPIE11823E..1WG}
{Gonzales}, M., {Stelter}, R.~D., {Kruglikov}, G., {Ratliff}, C., {Deich}, W.,
  {MacDonald}, N., {Sallum}, S., and {Skemer}, A., ``{Cryogenic test results of
  the SCALES focal plane coronagraph mechanism},'' in [{\em Society of
  Photo-Optical Instrumentation Engineers (SPIE) Conference
  Series}{\nolinebreak\hspace{0.1em}]},  {\em Society of Photo-Optical
  Instrumentation Engineers (SPIE) Conference Series} {\bf 11823},  118231W
  (Sept. 2021).

\bibitem{2014PASP..126..739R}
{Rauscher}, B.~J., {Boehm}, N., {Cagiano}, S., {Delo}, G.~S., {Foltz}, R.,
  {Greenhouse}, M.~A., {Hickey}, M., {Hill}, R.~J., {Kan}, E., {Lindler}, D.,
  {Mott}, D.~B., {Waczynski}, A., and {Wen}, Y., ``{New and Better Detectors
  for the JWST Near-Infrared Spectrograph},'' {\em PASP}~{\bf 126},  739 (Aug.
  2014).

\end{thebibliography}
\bibliographystyle{spiebib} 

\end{document}